\documentclass{aa}  

\usepackage[varg]{txfonts}

\usepackage{graphicx}	
\usepackage{amsmath}	
\usepackage{amssymb}	
\usepackage{amsfonts}

\usepackage{multicol}
\usepackage[normalem]{ulem}
\usepackage{bm}
\usepackage{color}
\usepackage{orcidlink}

\usepackage{hyperref}               
\bibpunct{(}{)}{;}{a}{}{,}             

\definecolor{grey}{rgb}{0.75,0.75,0.75}
\definecolor{Orange}{rgb}{1.0,0.5,0.15}
\definecolor{brown}{rgb}{0.7,0.25,0.0}
\definecolor{pink}{rgb}{1.0,0.5,0.5}
\definecolor{darkerred}{rgb}{0.8,0,0}
\definecolor{darkerblue}{rgb}{0,0,0.8}
\definecolor{Blue}{rgb}{0,0.08,0.65}
\definecolor{Red}{rgb}{0.65,0.08,0.05}
\definecolor{Green}{rgb}{0.15,0.45,0.25}

\newcommand{\rcm}{\bm{r}_{\mathrm{cm}}}
\newcommand{\qcm}{\bm{q}_{\mathrm{cm}}}
\newcommand{\cm}{\mathrm{cm}}

\newcommand{\dd}{\mathrm{d}}
\newcommand{\avg}[1]{\left\langle #1 \right\rangle}
\newcommand{\kk}{\bm{k}}
\newcommand{\rr}{\bm{r}}
\newcommand{\qq}{\bm{q}}

\newcommand{\vv}{\bm{v}}

\newcommand{\tr}{\mathrm{tr}}

\newcommand{\N}{\mathcal{N}}
\newcommand{\dirac}{\delta_\mathrm{D}}

\begin{document}

\title{The role of energy shear in the collapse of protohaloes}


\author{Marcello Musso\thanks{mmusso@usal.es} \inst{1} \orcidlink{0000-0003-1640-2795}
        \and
        Ravi K. Sheth\thanks{shethrk@upenn.edu} \inst{2,3} \orcidlink{0000-0002-2330-0917}
        }

\institute{Departamento de F\'{\i}sica Fundamental and IUFFyM,
            Universidad de Salamanca, Plaza de la Merced s/n, 37008 Salamanca, Spain
        \and
            Center for Particle Cosmology, University of Pennsylvania, 209 S. 33rd St., Philadelphia, PA 19104, USA
        \and
            The Abdus Salam International Center for Theoretical Physics, Strada Costiera 11, 34151 Trieste, Italy 
            }

\date{}


\abstract{Dark matter haloes form from the collapse of matter around special positions in the initial field, those where the local matter flows converge to a point. For such a triaxial collapse to take place, the energy shear tensor---the source of the evolution of the inertia tensor---must be positive definite. It has been shown that this is indeed the case for the energy shear tensor of the vast majority of protohaloes.  
At generic positions in a Gaussian random field, the trace and traceless parts of the tensor are independent of one another.  Here we show that, on the contrary, in positive definite matrices they correlate strongly, and these correlations are very similar to those exhibited by protohaloes.  Moreover, while positive-definiteness ensures that an object will collapse, it does not specify when.  Previous work has shown that the trace of the energy tensor---the energy overdensity---exhibits significant scatter in its values, but must lie above a critical `threshold' value for the halo to collapse by today. 
We show that suitable combinations of the eigenvalues of the traceless part are able to explain a substantial part of the scatter of the trace. These variables provide an efficient way to parameterise the initial value of the energy overdensity, allowing us to formulate an educated guess for the threshold of collapse. We validate our ansatz by measuring the distribution of several secondary properties of protohaloes, finding good agreement with our analytical predictions.
}


\maketitle

%

\allowdisplaybreaks

\section{Introduction}
The assembly, abundance and clustering of dark matter haloes play an important role both in galaxy formation and in the ongoing efforts to constrain cosmological parameters, where it is crucial to model accurately how haloes and galaxies trace the matter density field.  
While the latter is continuous, though, at least above some scale, the halo centers of mass define a point process at a given time, and the same is true at the initial time for the centers of mass of the protohalo patches from which they formed \cite{bbks86}.  Therefore, models of halo collapse and formation seek to identify what makes these locations special in the initial fluctuation field.  Such studies make two types of choices:  what is the `field', and what are the `properties' of this field?  

Following \cite{gg72}, who described halo formation as the collapse of a homogeneous sphere, most of the literature on the subject has assumed that the initial matter overdensity field is the one of interest.  However, haloes are not spherical. Thus, some early works suggested that the `shape tensor'---identified with the Hessian of the matter density field, that is the $3\times 3$ matrix of the second spatial gradients---provides useful additional information about departures from sphericity \citep{bbks86, vdwb96}.  

Other authors focused on a different $3\times 3$ matrix: the `deformation tensor'---constructed from the second gradients of the gravitational potential, averaged over the volume---noting that it is dynamically more interesting \citep{doroshkevich1970,bm96}.  Indeed, in the homogeneous ellipsoidal collapse model studied by \cite{bm96}, an initially spherical region is deformed by the surrounding shear field:  hence the name `deformation' tensor. (In fluid dynamics, the deformation tensor usually refers to infinitesimal volumes, not to the average over macroscopic regions.) The mean matter overdensity, thanks to Poisson's equation, is the trace of the deformation tensor; and since the trace is the sum of the three eigenvalues, this gives rise to the question:  do the 3 eigenvalues (or their linear combinations) themselves play an important role, or is only their sum important?  

Any $3\times 3$ matrix also has three `rotational invariants'. These are the coefficients of its characteristic polynomial, whose roots are the eigenvalues, and provide an alternative description of the same information. Since the trace is just one of the three, it is natural to ask what role the other two, which are {\em non}linear combinations of the eigenvalues, play.  This is particularly interesting in view of the fact that rotational invariants have been used to quantify how protoahaloes correlate with their environment \citep{Desjacques_local_2012,genLagbias,beyondLIMD}.  

Indeed, the second order invariant of the deformation tensor---the amplitude of the traceless shear---matters for halo formation:  protohaloes with larger shear tend to be more overdense \citep{smt01,dts13}, and this sources a correlation with the environment on larger scales 
\citep{scs13,ModiCastSel16,beyondLIMD}.  This is slightly surprising because, in a Gaussian random field, the two invariants are statistically independent \citep{st02}.  What is it about halo formation that induces a correlation between overdensity and shear?  Is this another manifestation of what makes protohalo positions special?
  
Recently, it has been highlighted that, as a `field', the initial energy overdensity is more interesting than the initial matter overdensity \citep{epeaks}.  Indeed, the energy shear tensor is more directly related to the physics of collapse than is the deformation tensor \citep{mds24}.  This implies that, unlike the deformation tensor, the energy shear is almost always positive definite.  Moreover, just as for the deformation tensor and even more strongly, the first and second invariants of the energy shear are positively correlated.  

With this in mind, our study has two main goals.  First, does the third invariant (of the energy shear tensor) also matter?  And second, can the correlations between the invariants of the protohalo patches be understood as simply arising from the positivity constraint?   

In the next section, we define the three invariants, before showing that all three play an important role in characterising the properties of protohalo patches.  Then we show that requiring positivity induces correlations between the invariants, which are qualitatively similar to those exhibited by protohaloes.  While positivity implies that an object will collapse, it does not determine when the collapse is complete:  Section~\ref{sec:+2crit} discusses how collapse by a specified time can be associated with a threshold in the energy overdensity. It then explores if this threshold depends on other invariants (than the trace) as well.   
A final section summarises our conclusions.

\section{Rotational invariants of protohaloes}\label{sec:inv}

The energy overdensity tensor of a generic comoving volume $V$, or energy shear, is the $3\times 3$ matrix
\begin{equation}
    u_{ij} \equiv \frac{3}{I}\int_V\dd\rr \rho(\rr,t)(r_i-r_{\cm,i})(\nabla_j\phi-[\nabla_j\phi]_\cm)
    \label{eq:uijdef}
\end{equation}
as defined by \cite{mds24}, where $\phi$ is the potential energy perturbation (related to the density contrast $\delta=\rho/\bar\rho-1$ by the Poisson equation $\nabla^2\phi=\delta$), $I\equiv\int_V\dd\rr\rho(\rr,t)|\rr-\rcm|^2$ is the moment of inertia, $\rcm$ the center of mass position and $[\bm{\nabla}\phi]_\cm$ its acceleration.

At very early times, when $\rho$ is nearly uniform and $\phi$ is a nearly Gaussian field, measuring $u_{ij}$ at generic positions returns a Gaussian random matrix. If $V$ is a randomly placed sphere of radius $R$, and only in this case, the covariance of  the energy shear can be expressed as
\begin{equation}
    \avg{u_{ij}u_{kl}} = \frac{\delta_{ik}\delta_{jl}+\delta_{il}\delta_{jk}+\delta_{ij}\delta_{kl}}{15}
    \sigma_{02}^2\,,
\end{equation}
where the Gaussian moment
\begin{equation}
    \sigma_{02}^2(R) \equiv \int\frac{\dd\kk}{(2\pi)^3}\delta(\kk)\frac{15j_2(kR)}{(kR)^2}
    \label{eq:sigma02}
\end{equation}
is the variance of the trace of $u_{ij}$, or energy overdensity \citep{epeaks}. All higher cumulants of the distribution vanish. Since $M=(4\pi/3)\bar\rho R^3$, $\sigma_{02}$ encodes the mass dependence of the distribution.

However, as pointed out recently by \cite{mds24}, $u_{ij}$ also sources the evolution of the inertia tensor $I_{ij}$ of the body. Therefore, if $V$ is a protohalo, the energy shear (more precisely, its symmetric part $u_{ij}+u_{ji}$) must have three positive eigenvalues for the volume to collapse along three axes. In other words, measuring $u_{ij}$ in protohaloes no longer yields a Gaussian random matrix, but a positive definite one.

\subsection{Rotational invariants of positive definite matrices}\label{sec:mat}

Regardles of its signature, the symmetric part $u_{ij}+u_{ji}$ of the energy shear has 3 real eigenvalues $\lambda_1>\lambda_2>\lambda_3$, a priori not necessarily positive, in terms of which one can construct the three rotational invariants
\begin{align}
    &\epsilon \equiv \sum_i \lambda_i, \qquad
     q^2\equiv \frac{3}{2}\sum_i (\lambda_i - \epsilon/3)^2, \quad {\rm and}\nonumber\\
    &U^3 \equiv \sum_i (\lambda_i - \epsilon/3)^3 
    = 3 \prod_i(\lambda_i - \epsilon/3)\,.
 \label{eq:eumu}
\end{align}
The trace $\epsilon = \tr(\mathbf{u})$ is the energy overdensity; in terms of $\bar u_{ij} \equiv u_{ij}-\epsilon\delta_{ij}/3$, the traceless part of the matrix, $q^2=3\tr(\mathbf{\bar u}^2)/2$ is the traceless shear amplitude; and $U^3=\tr(\mathbf{\bar u}^3)=3\det(\mathbf{\bar u}^3)$ is (thrice) the traceless determinant.  

\begin{figure}
    \centering
    \includegraphics[width=\linewidth]{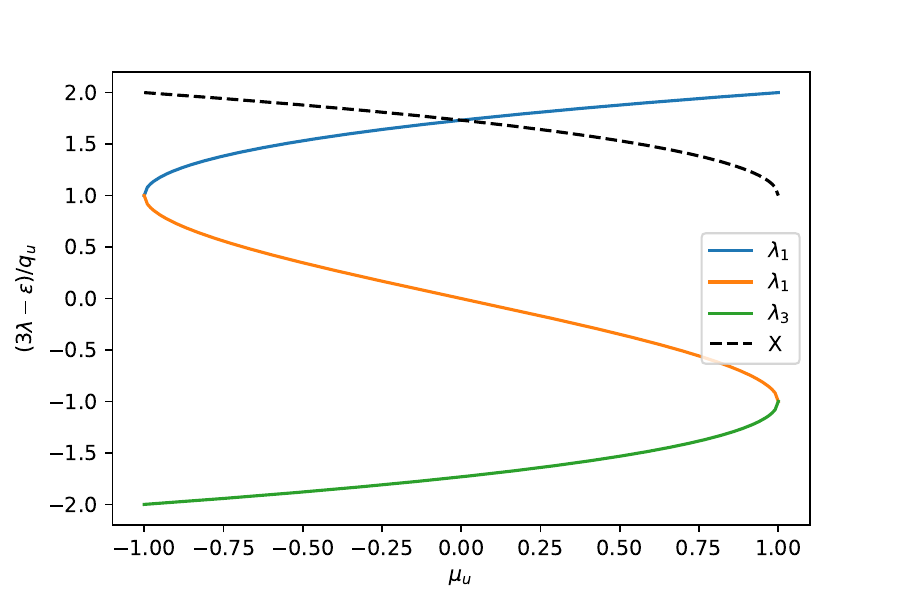}
    \caption{Ratio of traceless eigenvalues to $q/3$ as a function of $\mu$ (equation~\ref{eq:eig_roots}). Also shown (dashed line) is the function $X(\mu)$ defined in equation \eqref{eq:eps_constraint}, marking the lower limit of $\epsilon/q$}
\label{fig:lambdas}
\end{figure}

The equation for the eigenvalues can be conveniently written in the form of a depressed cubic as
\begin{equation}
    [(3\lambda-\epsilon)/q]^3 -3[(3\lambda-\epsilon)/q] - 2\mu = 0\,,
\label{eq:eig_eq}
\end{equation}
where
\begin{equation} 
  \mu\equiv (9/2)\,(U^3/q^3)\,;
\end{equation}
for this equation to admit three real solutions, as it should be since the matrix is symmetric, the constraint $\mu^2<1$ must be satisfied, or equivalently $-1\leq\mu\leq1$. Clearly, one also has $q \equiv \sqrt{q^2}\ge 0$. The roots of the cubic equation are then
\begin{equation}
    \lambda_k = \frac{\epsilon}{3} +\frac{2q}{3}
    \cos\bigg[\frac{1}{3}\arccos(\mu)-\frac{2\pi}{3}(k-1)
    \bigg]\,,
\label{eq:eig_roots}
\end{equation}
which expresses the three eigenvalues in terms of the three rotational invariants.
Configurations with $\mu\simeq1$ have $\lambda_1>\lambda_2\simeq\lambda_3$, whereas 
$\mu\simeq-1$ implies $\lambda_1\simeq\lambda_2>\lambda_3$, and $\mu\simeq 0$ implies $\lambda_2\simeq \epsilon/3$.
Figure \ref{fig:lambdas} shows the three eigenvalues as a function of $\mu$.  Evidently, 
  $3(\lambda_1-\lambda_3)/q \approx 3$ 
for all values of $\mu$, meaning that $\lambda_1-\lambda_3\approx q$. Hence, the ellipticity $e_u\equiv (\lambda_1-\lambda_3)/2\epsilon\approx q/2\epsilon$.  

Regardless of the sign of the eigenvalues, the rotational invariants always satisfy the relations
\begin{align}
  &\epsilon^2-q^2 = 3\,(\lambda_1\lambda_2 + \lambda_1\lambda_3 + \lambda_2\lambda_3) \quad{\rm and} \label{eq:eq}\\
  &2\mu + \frac{\epsilon}{q}\bigg(\frac{\epsilon^2}{q^2}-3\bigg)  = 
  27\frac{\det(\mathbf{u})}{q^3} 
  =27\frac{\lambda_1\lambda_2\lambda_3}{q^3}\,. \label{eq:mu}
\end{align}
If however $u_{ij}$ is positive definite (that is, if all its eigenvalues are positive), 
the right hand sides of equations \eqref{eq:eq} and \eqref{eq:mu} are both positive; so $\epsilon>q$ and $(\epsilon/q)^3-3(\epsilon/q)+2\mu>0$. 
The equation associated to the latter inequality is the same as the equation for $(\epsilon-3\lambda)/q$, and its roots can be obtained from equation \eqref{eq:eig_roots}. 

The only root larger than 1 (required since $\epsilon>q$) is the one obtained plugging in $\lambda_3$. Therefore, all eigenvalues are positive if (and only if)
\begin{equation}
    \frac{\epsilon}{q} > 
    - 2\cos\bigg[\frac{\arccos(\mu)-4\pi}{3}\bigg] \equiv X(\mu)
\label{eq:eps_constraint}
\end{equation}
(which also implies that $\epsilon > q$). Clearly, this is the same as requesting that $\lambda_3>0$, since $X(\mu)$ is equal to $(\epsilon-3\lambda_3)/q$. Using trigonometric identities, $X(\mu)$ can also be expressed as
\begin{equation}
    X(\mu)=2\cos\bigg[\frac{\arccos(-\mu)}{3}\bigg]\,, \quad \mathrm{with}\quad 1\leq X\leq 2\,. 
    \label{eq:X}
\end{equation}
Its numerical value is also shown in Fig.~\ref{fig:lambdas} (dashed line).

Although we have focused on the rotational invariants, other combinations of the eigenvalues have featured in previous work \citep{bbks86, bm96, sandvik_vw2007}.  Quantities built from linear combinations of the eigenvalues tend to all have the same structure:  one variable is the trace, which we denote $\epsilon$, and the two others are formed from differences with respect to one of the eigenvalues (that is, they are constructed from the traceless matrix). E.g., 
\begin{equation}
    v_\pm = (\lambda_1-\lambda_3) \pm (\lambda_2-\lambda_3),
    \label{eq:vpm}
\end{equation}
which obviously has $v_\pm\ge 0$ and $0\le v_-\le v_+$.  
The shape parameters $y$ and $z$ defined in equation~(A2) of \cite{bbks86} are formed from differences of the other eigenvalues with respect to $\lambda_2$ rather than $\lambda_3$.  (The ellipticity and prolateness of \cite{bm96} are conceptually the same:  differences with respect to the middle eigenvalue.)
Our $v_\pm$ variables were considered by \cite{sandvik_vw2007}, though they had no real physical justification for their choice.  Likewise, one could have worked with differences with respect to $\lambda_1$.  We prefer differences with respect to $\lambda_3$ because the positivity constraint ($\lambda_3\ge 0$) is particularly simple in these variables.  This is because 
\begin{align}
    v_+ &= \lambda_1 + \lambda_2 - 2\lambda_3 
         = \epsilon-3\lambda_3 = qX(\mu)
  \label{eq:vqX}
\end{align}
so positivity requires $\epsilon\ge 0$ and $0\le v_+\le \epsilon$ but the $v_-$ range is unchanged ($0\le v_-\le v_+$).
As we will see below, this makes it easy to describe the joint distribution of $(\epsilon,v_+,v_-)$ when all three eigenvalues are positive. 

\subsection{Protohalo properties}
We study the properties of protohaloes identified using a Spherical Overdensity halo finder using a threshold of $319$ times the background density in the $z=0$ outputs of the Bice and Flora simulations of the SBARBINE simulation suite \citep{despali16}.  For both simulations, the 
background cosmology has $\Omega_m = 0.307$, $\Omega_\Lambda = 0.693$, $\sigma_8 = 0.829$ and $h = 0.677$.  
Each simulation contains $1024^3$ dark matter particles in a periodic box of side $L_{\mathrm{box}}=125h^{-1}$Mpc (Bice) and $L_{\mathrm{box}}=2h^{-1}$Gpc (Flora), so the corresponding particle masses are $1.55\times10^{8} h^{-1}M_\odot$ for Bice and $6.35\times10^{11} h^{-1}M_\odot$ for Flora.

\begin{figure}
    \centering
    \includegraphics[width=\columnwidth]{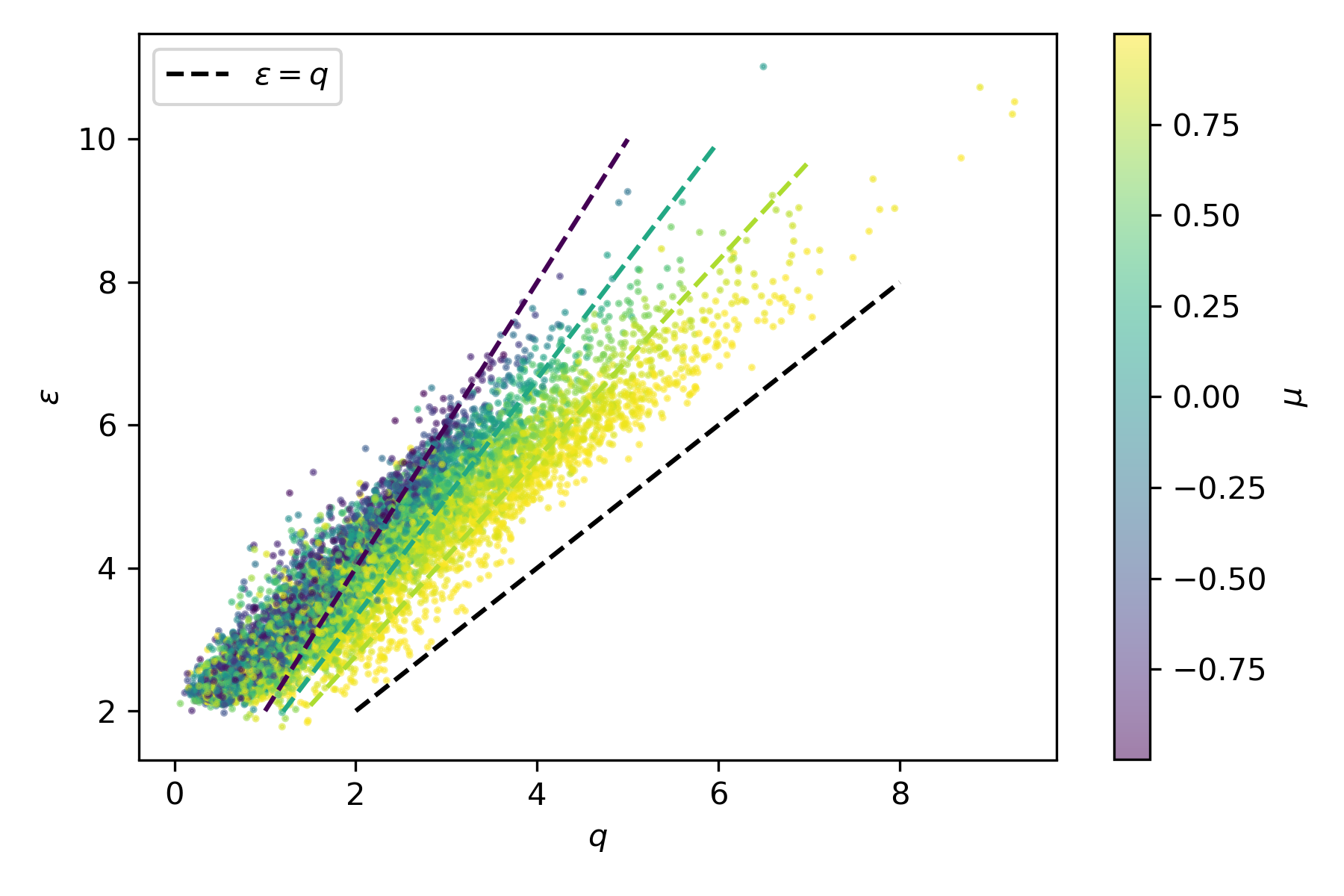}
    \caption{Correlation between the three rotational invariants of protohaloes:  energy overdensity $\epsilon$, the traceless energy shear $q$, and $\mu$ (see equation~\ref{eq:eumu}). The color coding displays the value of $\mu$.  Dashed lines show $\epsilon=Xq$ with $X = n/3$ for $n=[3,4,5,6]$, corresponding to $\mu=[1, 0.815, 0.185, -1]$ (from bottom to top).}
    \label{fig:eps_vs_u}
\end{figure}

Our Bice sample contains 5000 randomly selected haloes with masses between $10^{11}$ and $4\times 10^{13}h^{-1}M_\odot$.  
Our Flora sample contains haloes from three separate mass bins, as follows: 2000 randomly selected haloes with masses between $4\times 10^{13}$ and $10^{14}h^{-1}M_\odot$, 
2000 randomly selected haloes with masses between $10^{14}$ and $10^{15}h^{-1}M_\odot$, and all 1378 haloes more massive than $10^{15}h^{-1}M_\odot$.  In the following, we will indicate these three samples as Flora-S, Flora-M and Flora-L respectively.

For each halo identified at $z=0$, we use `protohalo' to refer to the region occupied by that halo's particles in the initial conditions.  Following \cite{mds24}, we measured each protohalo's center of mass position and velocity, $\qcm$ and $\bm{v}_\cm$, by averaging over all its particles. We then estimated its energy shear tensor as:
\begin{align}
  \hat u_{ij} &\equiv -3\, \frac{\sum_{n=1}^{N_H} (\qq^{(n)}-\qcm)_i(\bm{v}^{(n)} - \bm{v}_\cm)_j/fDH}{
  \sum_{n=1}^{N_H} |\qq^{(n)}-\qcm|^2} \,,
\label{eq:esteps}
\end{align}
where $D$ is the $\Lambda$CDM density perturbation growth factor (at the redshift $z$ of the snapshot), $f=\dd\ln D/\dd\ln a$, and $n$ runs over the $N_H$ protohalo particles. 
(Note that $u_{ij}$ is a `dimensionless energy' because, in the initial conditions, the velocity is proportional to the acceleration which is the gradient of the gravitational potential: $\nabla\phi \simeq \vv/\dot D= \vv/fDH$.)  We then diagonalize $\hat u_{ij}$ and estimate $\epsilon$, $q$ and $\mu$ using Eq.~(\ref{eq:eumu}).
\begin{figure}
    \centering
    \includegraphics[width=\columnwidth]{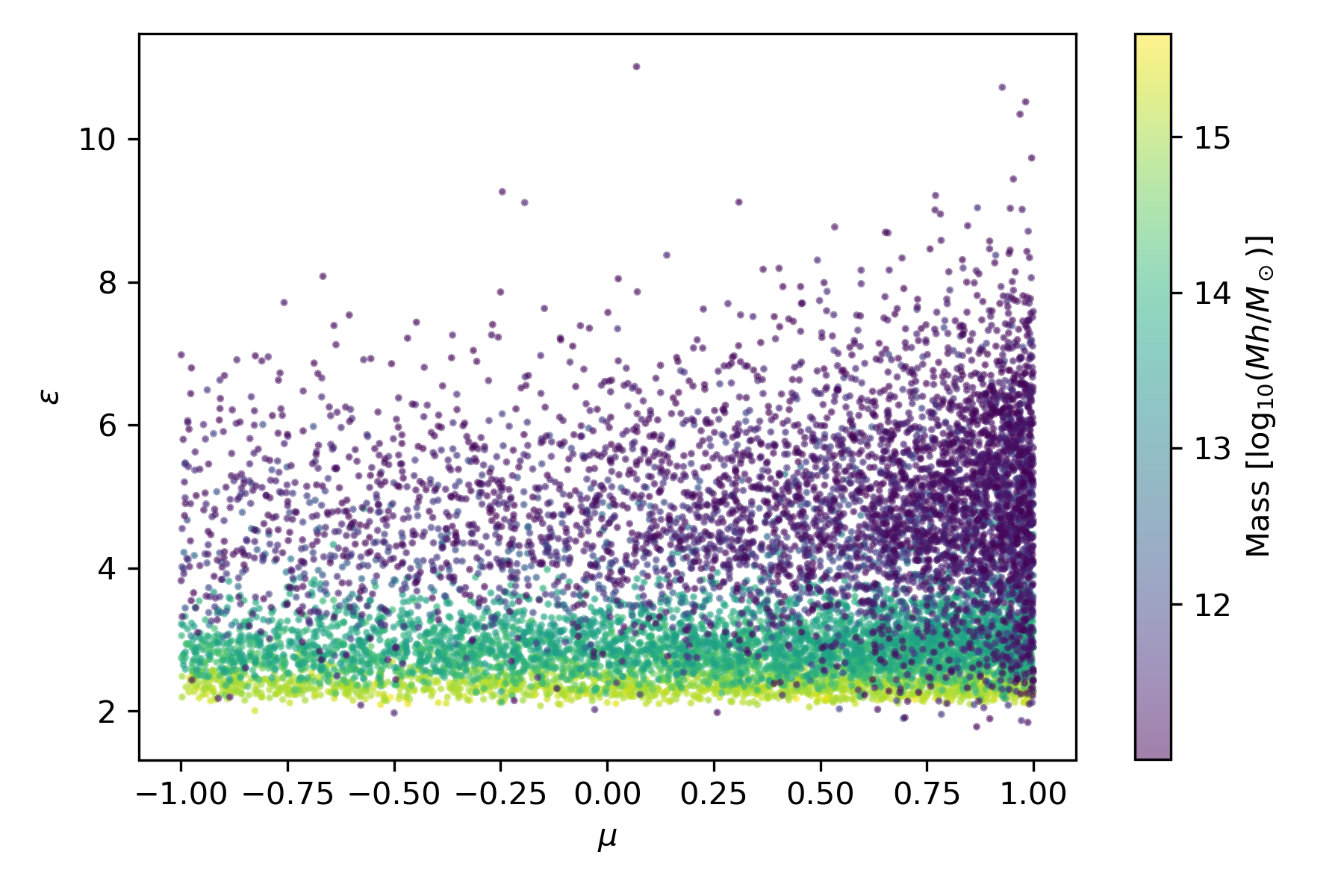}
    \includegraphics[width=\columnwidth]{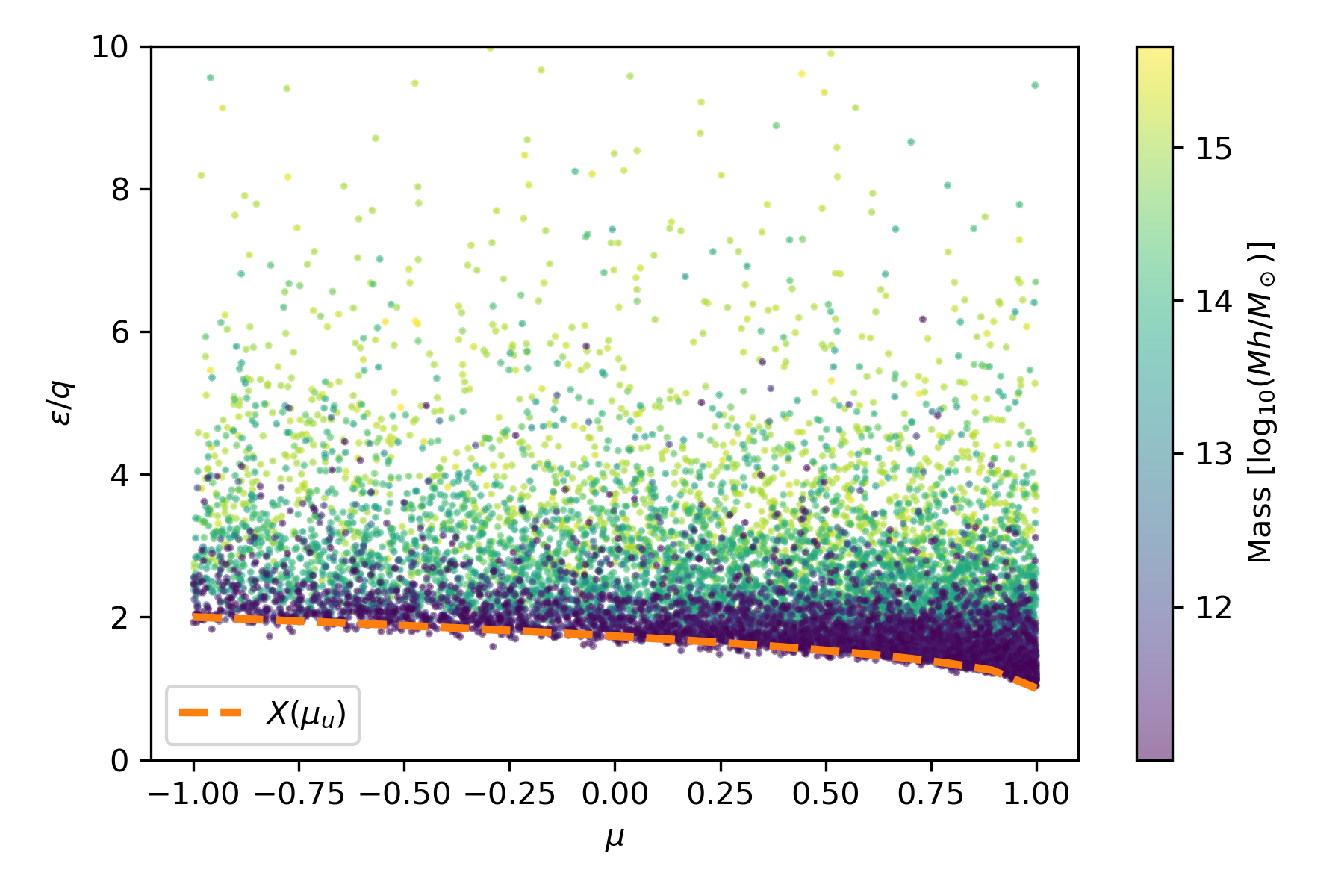}
    \caption{Correlation of $\epsilon$ (top panel) and $\epsilon/q$ (bottom) with $\mu$ for protohaloes, color coded by $\log_{10}(M/h^{-1}M_\odot)$. The dashed line in the bottom panel shows the lower limit given by equation \eqref{eq:eps_constraint}.}
    \label{fig:eps_over_mu}
\end{figure}

Figure~\ref{fig:eps_vs_u} shows a scatter plot of $\epsilon$ vs $q$, with the color indicating the value of $\mu$, measured in the same set of protohaloes that was used in \cite{mds24}, where the underlying cosmology and halo identification schemes are described.  Indeed, the points here are the same as in their Figure~10; the {\em only} difference is that we have colored each point by its (protohalo's) value of $\mu$.  Clearly, the three invariants are strongly correlated:  larger shear has larger $\epsilon$, with the slope of the (nearly linear) scaling being determined by the third invariant (points with larger $\mu$ have $\epsilon$--$q$ relations that are shifted down and/or to the right).  This is remarkable because for generic Gaussian matrices, and thus for tensors measured at generic positions in a Gaussian random field, any variables constructed from the traceless part of the matrix should be {\em un}correlated with the trace.  

This correlation provides yet another indication that protohaloes are special locations in the initial density field. As a precursor of one of our results, consider what happens if the energy shear is positive definite, as \cite{mds24} showed it must be for protohalo patches. Then, as discussed in Section \ref{sec:mat}, one must have that $\epsilon>q$.  This introduces the linear correlation seen in Fig. \ref{fig:eps_vs_u}. Secondly, equation \eqref{eq:eps_constraint} must be satisfied, and, since its r.h.s. is a decreasing function of $\mu$, protohaloes with $\mu$ closer to $-1$ are constrained to have larger $\epsilon/q$.
This explains why, as one moves away from the lower dashed line in Fig.~\ref{fig:eps_vs_u} (indicating where $\epsilon=q$), the value of $\mu$ tends to decrease. The three steeper dashed lines indicate the lower limit for haloes with $\epsilon/q$ larger than 4/3, 5/3 and 2 respectively (corresponding to $\mu$ of .185, .815 and 1). Moreover, as shown by equation \eqref{eq:eq}, for a positive definite matrix the inequality $\epsilon>q$ is saturated only if all three eigenvalues simultaneously vanish, which never happens in practice.

Figure \ref{fig:eps_over_mu} (top panel) shows the values of $\epsilon$ versus the third rotational invariant $\mu$, coloured by the logarithm of the mass. The yellow and green bands at the bottom correspond to the Flora-L, Flora-M and Flora-S samples, and are quite uniformly distributed between -1 and 1. Conversely, Bice haloes (lower mass) correspond to the violet dots of the upper band, and show a clear tendency to prefer values closer to 1. This behaviour can also be understood from equation \eqref{eq:eps_constraint}, since for equal values of $q$ protohaloes with $\mu\simeq-1$ must have twice as large values of $\epsilon$ as those with $\mu\simeq1$, and are therefore less likely. Furthermore, apart from a few low-mass outliers, in all protohaloes $\epsilon$  is larger than roughly 2.2 (we will denote this bottom value $\epsilon_c$ in the following), and tends to be higher at smaller masses. The trend of $\epsilon$ with mass and $q$ are mutually compatible, since large-mass haloes prefer small values of $q$ and thefore lie in the bottom left corner of Fig. \ref{fig:eps_vs_u}.

The bottom panel of Fig. \ref{fig:eps_over_mu} displays instead the value of the ratio $\epsilon/q$, and clearly shows the lower cut determined by equation \eqref{eq:eps_constraint} (dashed line in the plot). This bound affects the low-mass Bice haloes the most. Large-mass protohaloes have small $q$, but have nevertheless (as the top panel shows) nonzero $\epsilon$. Therefore, their $\epsilon/q$ is naturally large, and mostly unaffected by the presence of a lower bound.

\begin{figure}
    \centering
    \includegraphics[width=\columnwidth]{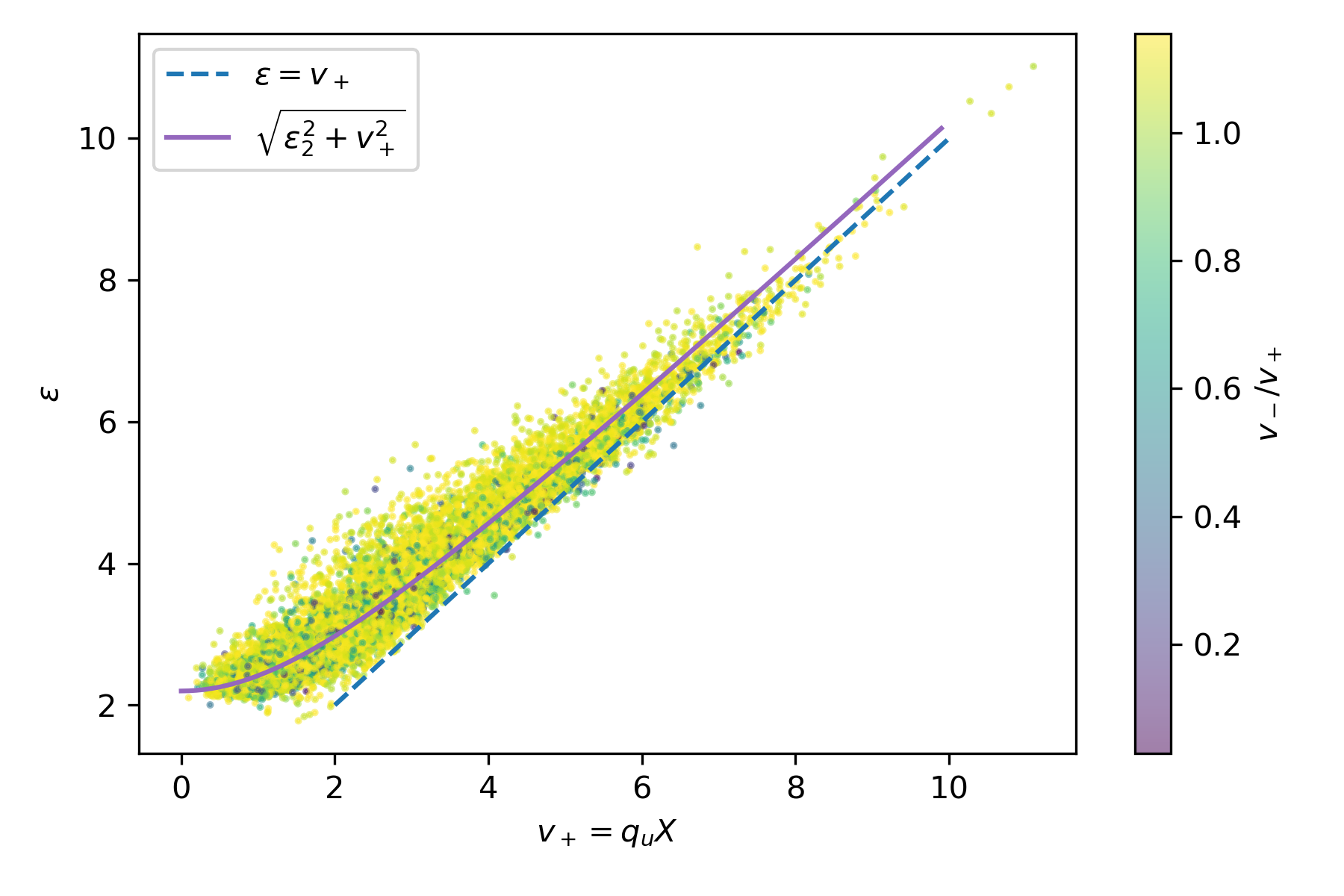}
    \caption{Correlation of $\epsilon$ with $v_+$, with colors showing the value of $v_-/v_+$. The scatter is significantly narrower than in Figure \ref{fig:eps_vs_u}, especially at larger values, and there is no residual correlation signal with $v_-/v_+$. Dashed line shows the theoretical lower limit $\epsilon=v_+$; solid line shows  $\sqrt{\epsilon_c^2+v_+^2}$ with $\epsilon_c=2.2$.}
    \label{fig:eps_vs_uX}
\end{figure}

The fact that the correlation pattern is explained by the $\epsilon > qX(\mu)$ constraint motivates us to look at the correlation structure with the variables $v_+= qX$ and $v_-$ defined in equation \eqref{eq:vpm}. As shown in Figure \ref{fig:eps_vs_uX}, the linear correlation between $\epsilon$ and $v_+$ is even more prominent, with a scatter that becomes significantly narrower at large values. Points in the graph are colored by the value of $v_-/v_+$ in each protohalo: the fact that colors are uniformly distributed in the plot shows that there is no residual correlation of $\epsilon-v_+$ with $v_-/v_+$.

\subsection{Distribution of rotational invariants of positive definite matrices}

To simplify notation, in this section and in the following ones we set $\sigma_{02}=1$, unless otherwise specified. The distributions of $\epsilon$ and $q$ presented here should therefore be read as those of $\epsilon/\sigma_{02}$ and $q/\sigma_{02}$. Being defined as a ratio of variables, $\mu$ is not affected by this choice.

\cite{doroshkevich1970} shows that the joint distribution of the 3 eigenvalues of a $3\times 3$ symmetric Gaussian matrix is given by 
\begin{equation}
    p(\lambda_1,\lambda_2,\lambda_3) = \frac{3^35^3}{8\pi\sqrt{5}}{\rm e}^{-\epsilon^2/2}{\rm e}^{-5q^2/2}\,\Delta(\lambda_1,\lambda_2,\lambda_3),
    \label{eq:pl1l2l3}
\end{equation}
where $\epsilon$ and $q^2$ were defined earlier and 
\begin{equation}
    \Delta \equiv (\lambda_1-\lambda_2)(\lambda_1-\lambda_3)(\lambda_2-\lambda_3)
\end{equation}
is the Vandermonde determinant.  In terms of the rotational invariants $\epsilon$, $q$ and $\mu$ this distribution is 
\begin{equation}
    p(\lambda_1,\lambda_2,\lambda_3)
    \,\dd\lambda_1 \dd\lambda_2 \dd\lambda_3 
    = \N(\epsilon)\,\chi(q)\frac{\Theta(1-\mu^2)}{2}\,\dd\epsilon\dd q\dd\mu,
\end{equation}
where 
\begin{equation}
    \N(\epsilon)\,\dd\epsilon = 
    \frac{{\rm e}^{-\epsilon^2/2}}{\sqrt{2\pi}} \dd\epsilon
    \label{eq:gaussian}
\end{equation}
and
\begin{equation}
    \chi(q)\,dq = \frac{dq^2}{q^2}\,\left(\frac{5q^2}{2}\right)^{5/2} 
    \frac{{\rm e}^{-5q^2/2}}{\Gamma(5/2)}\,.
    \label{eq:chi5}
\end{equation}
Therefore, $\epsilon$ is Gaussian distributed, $5q^2$ follows a chi-square distribution with 5 degrees of freedom, and $\mu$ is uniform over $[-1, 1]$.  
In particular, note that the distributions of $\epsilon$, $q$ and $\mu$ are independent, so the joint distribution above cannot explain the correlations seen in Figure~\ref{fig:eps_vs_u}.

How is this modified if all three eigenvalues have the same sign?
If we require $\lambda_3>0$, then $q<\epsilon$ and $2\mu>(\epsilon/q)(3-\epsilon^2/q^2)$ 
In turn, this requires that ${\rm max}[-1,(\epsilon/2q)(3-\epsilon^2/q^2)]<\mu<1$.
So, integrating over $\mu$ returns 1 if $q<\epsilon/2$, but $[(\epsilon/2q)(\epsilon^2/q^2 - 3)+1]/2$ if $\epsilon/2<q<\epsilon$. The latter factors as $(\epsilon/q+2)(\epsilon/q-1)^2/2$. Hence, the probability of having all positive eigenvalues is 
\begin{align}
    p(+) &\equiv p(\lambda_3\ge 0)  \notag \\
    &= \int_0^\infty \dd\epsilon\N(\epsilon) \left[
    \int_0^{\epsilon/2} \dd q\,\chi(q) \right. \nonumber\\
    &\qquad\qquad\quad + 
    \left.\int_{\epsilon/2}^{\epsilon} \dd q\,\chi(q)\, 
    \frac{(\epsilon/q+2)(\epsilon/q - 1)^2}{2}\right] \nonumber\\
    &= \frac{1}{2} - \frac{3\arctan(\sqrt{5}) + 2\sqrt{5}}{6\pi} 
    \approx \frac{2}{25} \,.
     \label{eq:p+}
\end{align}
With this normalisation, the conditional distributions are
\begin{align}    
    p(\epsilon|+) &= \frac{\N(\epsilon)}{p(+)}\left[\int_0^{\epsilon/2} \dd q\,\chi(q)\right. \nonumber\\
    &\qquad + \left.\int_{\epsilon/2}^{\epsilon} \dd q\,\chi(q)\,\frac{(\epsilon/q+2)(\epsilon/q - 1)^2}{2}\right]\nonumber\\
    &= \frac{\N(\epsilon)}{2p(+)}\,\bigg[{\rm erf}\left(\frac{\epsilon}{2}\sqrt{\frac{5}{2}}\right) + {\rm erf}\left(\epsilon\sqrt{\frac{5}{2}}\right) \nonumber\\
    &\qquad - 3\sqrt{5}\,\epsilon\, \frac{{\rm e}^{-5\epsilon^2/8}}{\sqrt{2\pi}}\bigg]\,, \label{eq:pdfe+}\\
    p(q|+) &= \frac{\chi(q)}{p(+)} \left[
    \int_{2q}^{\infty} \dd\epsilon\,\N(\epsilon)\right. \nonumber\\
    &\quad\quad +  \left.\int_{q}^{2q} \dd\epsilon\,\N(\epsilon)\,\frac{(\epsilon/q+2)(\epsilon/q - 1)^2}{2}\right] \nonumber\\
    &= \frac{\chi(q)}{4p(+)}\bigg[{\rm erfc}(q/\sqrt{2}) + {\rm erfc}(q\sqrt{2}) \nonumber\\
    &\qquad\qquad - \frac{{\rm e}^{-2q^2}(q^2 + 2) + 2{\rm e}^{-q^2/2}(q^2 - 1)}{q^3\sqrt{2\pi}} \,
                 \bigg]\,, 
                 \label{eq:pdfq+}\\
    p(\mu|+) &=  \frac{1}{2p(+)} \int_0^{\infty}\dd q\, \chi(q)\int_{X(\mu)q}^\infty\!\!\dd\epsilon\,\N(\epsilon) \notag \\
    & = \frac{1}{2\pi p(+)}
    \bigg[\mathrm{arccot}\left(\frac{X}{\sqrt{5}}\right) - \frac{\sqrt{5}X(3X^2+25)}{3(X^2+5)^2}\bigg] \,, 
    \label{eq:pdfmu+}
\end{align}
where $X(\mu)$ was defined in equation \eqref{eq:X}.

This shows that requiring positive definiteness shifts the distribution of $\epsilon$ so that large values are more likely (e.g. its mean is non-zero, close to 1.65); it shifts the distribution of $q$ so that large values are less likely; it induces a correlation between $\epsilon$ and $q$; it changes the distribution of $\mu$ so that values of unity are more likely. For instance, some of the relevant moments are:
\begin{align}
    &\langle\epsilon|+\rangle \simeq 1.665\approx 5/3\,,\qquad \langle\epsilon^2|+\rangle \simeq 3.067\,,\nonumber\\
    &\langle u|+\rangle \simeq 0.725 \approx 29/40\,,\qquad \langle u^2|+\rangle\nonumber \simeq 0.587\,.
\end{align}
The cross-correlation between $\epsilon$ and $q$ is
$C_{\epsilon u} = \langle\epsilon u|+\rangle \simeq 0.068$.
The expected slope of the $\epsilon$--$q$ relation is then $C_{\epsilon u}/{\rm Var}(u|+) \simeq 0.068/0.063\approx 1$, in agreement with Figure~\ref{fig:eps_vs_uX}.

\begin{figure}
    \centering
    \includegraphics[width=\columnwidth]{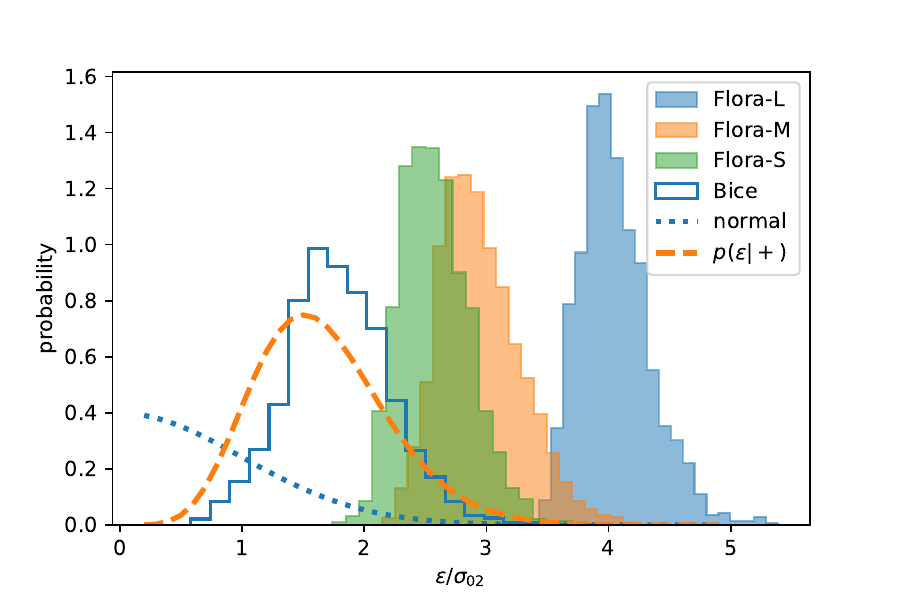}
    \caption{Distribution of trace $\epsilon/\sigma_{02}$ for a few ranges in halo mass (histograms).  Dotted curve shows the expected (zero-mean, unit variance Gaussian) distribution for unconstrained positions; dashed curve is for positions that have $\lambda_3\ge 0$ (Eq.~\ref{eq:pdfe+}).  More massive haloes have larger $\epsilon/\sigma_{02}$, but the distribution around the mean is approximately independent of mass.}
    \label{fig:hist_eps}
\end{figure}
\begin{figure}
    \centering
    \includegraphics[width=\columnwidth]{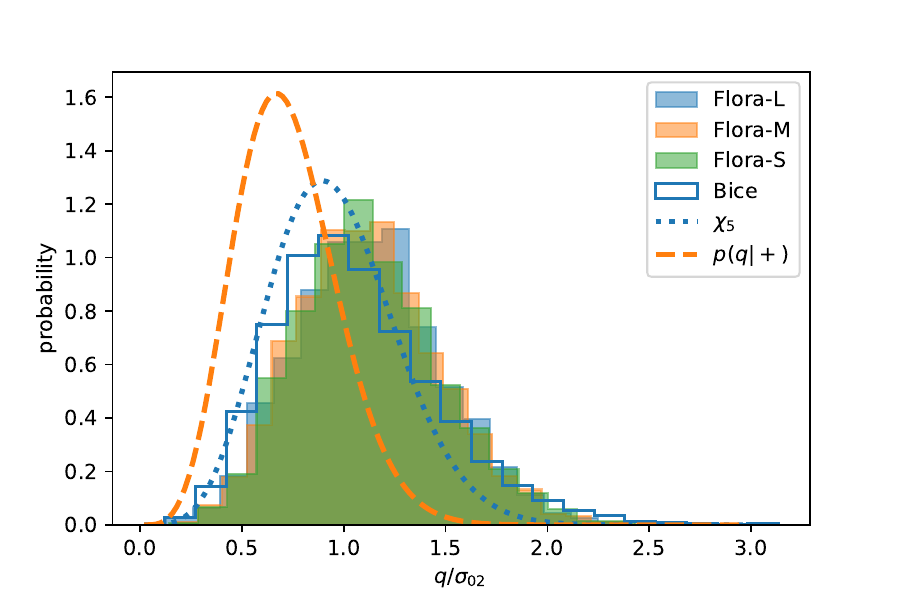}
    \includegraphics[width=\columnwidth]{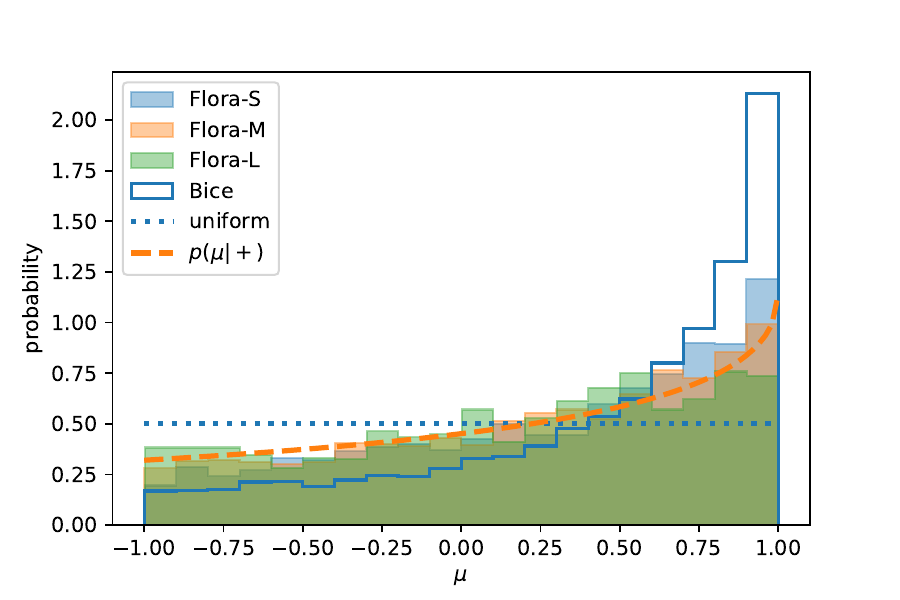}
    \caption{Distribution of traceless shear $q/\sigma_{02}$ (top) and $\mu$ (bottom). 
    Dotted curves show the expected distributions for unconstrained positions; dashed for positions that have $\lambda_3\ge 0$ (Eqs.~\ref{eq:pdfq+} and~\ref{eq:pdfmu+}).   Whereas $q/\sigma_{02}$ is approximately the same for all masses, $\mu$ is closer to uniform as mass increases.} 
    \label{fig:hists}
\end{figure}

These distributions of $\epsilon$, $q$ and $\mu$ are shown by the dashed lines in Figs. \ref{fig:hist_eps} and \ref{fig:hists} (dotted lines correspond to the standard results for Gaussian matrices with unconstrained eigenvalues: normal, chi-5 and uniform respectively). In each figure, the histograms correspond to the measurement of the same quantity in four halo samples (three filled histograms show Flora haloes; open histogram shows the less massive Bice sample). Figure \ref{fig:hist_eps} shows that, for large masses (filled histograms), the distribution of $\epsilon/\sigma_{02}$ is strongly mass dependent and is not well described by Equation \eqref{eq:pdfe+}. However, for small masses (leftmost histogram) the measurement tends to the predicted (mass-independent) result. This trend can be attributed to the fact that massive protohaloes have $\epsilon \gtrsim 2$ but $q\sim\sigma_{02}\ll2$, so they are almost insensitive to the positivity constraint $\epsilon>q$. However, most lower mass protohaloes have $q>2$, so for them $\epsilon>q$ is the dominant constraint, and this constraint explains most of their behaviour.

The top panel of Figure \ref{fig:hists} displays the distribution of the traceless shear amplitude $q/\sigma_{02}$.  In contrast to $\epsilon/\sigma_{02}$, this shear shows remarkably little mass dependence: the four histograms lie almost on top of each other. Nevertheless, the measured values tend to be larger than predicted by $p(q|+)$ (equation \ref{eq:pdfq+}), shown by the dashed curve.
The bottom panel shows the distribution of $\mu$. In this case, values of $\mu$ closer to 1 become more likely, and the trend is more pronounced for less massive haloes. This is a consequence of the constraint $X(\mu)<\epsilon/q$: since small haloes have $\epsilon/q\sim 1$, values of $\mu$ closer to 1 are favoured, for which $X(\mu)$ is smaller (see bottom panel of Fig. \ref{fig:eps_over_mu}). On the other hand, large haloes typically have $\epsilon/q\gg 1$, which makes the constraint less relevant (since $1<X(\mu)<2$). The same can also be inferred from Fig. \ref{fig:eps_vs_u}: more and more haloes are excluded as the dashed line becomes steeper, but this does not affect the bottom left corner (where most massive haloes lie).

\subsection{Linear combinations of eigenvalues}
Figure~\ref{fig:eps_vs_uX} shows that the $\epsilon$--$v_+$ relation has smaller scatter than $\epsilon$--$q$.  This motivates a study of the joint distribution of $(\epsilon, v_+, v_-)$ and how it is impacted by the positivity constraint.

Since $v_\pm$ are formed from differences of eigenvalues, we expect them to be independent of the trace.  Indeed, their joint distribution is (always with the $\sigma_{02}=1$ convention)
\begin{equation}
    p(\epsilon,v_+,v_-) = \frac{15^3}{2^5}\frac{{\rm e}^{-\epsilon^2/2}}{\sqrt{2\pi}}
    \frac{{\rm e}^{- 5v_+^2/8 - 15v_-^2/8}}{\sqrt{90\pi}}\,v_-\,(v_+^2-v_-^2) \,, 
    \label{eq:joint_evpvm}
\end{equation}
where an unconstrained Gaussian random field has 
$-\infty\le\epsilon\le\infty$, $v_+\ge 0$ and $0\le v_-\le v_+$.   Integrating over $v_-$ yields 
\begin{equation}
    p(\epsilon,v_+) = \N(\epsilon)\,
    \frac{{\rm e}^{- 5v_+^2/8}}{\sqrt{2\pi/5}}\left({\rm e}^{-15v_+^2/8} -1 + 15v_+^2/8\right)\,.
\label{eq:joint_eps_v+}
\end{equation}
It is straightforward to check that integrating $v_+$ over the full range ($v_+\ge 0$) yields $\N(\epsilon)$; and that, recalling that the positivity constraint is simply $0\le v_+\le \epsilon$ (with the $v_-$ range unchanged), integrating over this restricted range returns $p(\epsilon|+)\,p(+)$.  The distribution of $v_+$ subject to the constraint that $\lambda_3\ge 0$ is given by 
\begin{align}
    p(v_+|+) &= \int_{v_+}^\infty d\epsilon\, \frac{p(\epsilon,v_+)}{p(+)} \nonumber\\
    &= \frac{{\rm e}^{- 5v_+^2/8}}{\sqrt{2\pi/5}}
     \left({\rm e}^{-15v_+^2/8} -1 + 15v_+^2/8\right)\,
     \,\frac{{\rm erfc}(v_+/\sqrt{2})}{2p(+)}.
\end{align}
The final error function modifies the shape, effectively reducing the probability of large $v_+$.

\section{From collapsing to collapsed}\label{sec:+2crit}
The positive definite constraint is sensible from a dynamical point of view:  bound objects should collapse along all three axes.  
However, Fig.~\ref{fig:eps_vs_u} shows a clear lower limit, $\epsilon\gtrsim 2$, that (for small $q$) is more stringent than the $\epsilon>q$ limit required by positive definiteness (lowest dashed line).  This strongly suggests that something in addition to positive definiteness is required to explain the properties of protohaloes (in this case, the patches destined to be haloes by the present time, $z=0$). Simply requiring $\lambda_3\ge 0$ does not constrain {\em when} the collapse is completed. There may be many objects that are collapsing now but have not yet collapsed completely, and others that collapsed earlier.  This raises the question of what sets the time of collapse.  

As shown by \cite{epeaks}, $\epsilon$ controls the evolution of the inertial radius of the halo; for a centrally peaked density profile, $\epsilon$ is always larger than mean enclosed matter overdensity, which controls the evolution of the geometric radius (related to the cubic root of the volume). In a simple spherical collapse model, the latter must have an initial overdensity of $\sim 1.7$ for the required density to be assembled today \citep{gg72}. Since the final inertial radius is always smaller than the geometric radius, the threshold on $\epsilon$ must be even larger.  \cite{epeaks} showed that in large haloes, whose collapse should be fairly anisotropic, $\epsilon$ must equal a critical value: $\epsilon_c\approx 2$ .

A simple calculation of the distribution of the three rotational invariants $\epsilon$, $q$ and $\mu$ subject to the additional constraint that $\epsilon>\epsilon_c$ is given in Appendix \ref{sec:app_ec}. However, the conditional pdf for $\epsilon$ resulting from this approach is simply the $\epsilon>\epsilon_c$ tail of equation \eqref{eq:pdfe+}, appropriately normalized.  None of our measurements is close to such a sharply cut distribution, so we now explore a different path.

In the current context, it is natural to assume that the other rotational invariants also determine the collapse time. Requesting collapse by today will thus enforce a relation 
\begin{equation}
    \mathcal{C}(\epsilon, q, \mu, \epsilon_c) = 0,
\end{equation}
where $\mathcal{C}$ is a function of the three invariants and of a constant $\epsilon_c$ that depends on the collapse time.  
For example, a simple perturbation to the spherical collapse model might be written as 
$\epsilon \simeq \epsilon_c +q$, 
with $\epsilon_c\approx 2$.  Indeed, in models based on density rather than energy, it has been proposed that $\delta - \delta_c \propto q$ \cite[e.g.][]{scs13,lud_borz_porc14,Castorina:2016tuc}.  This cannot be the full story, though, because Fig.~\ref{fig:eps_vs_u} shows that there is substantial scatter in the $\epsilon$--$q$ plane.  

\begin{figure}
    \centering
    \includegraphics[width=\linewidth]{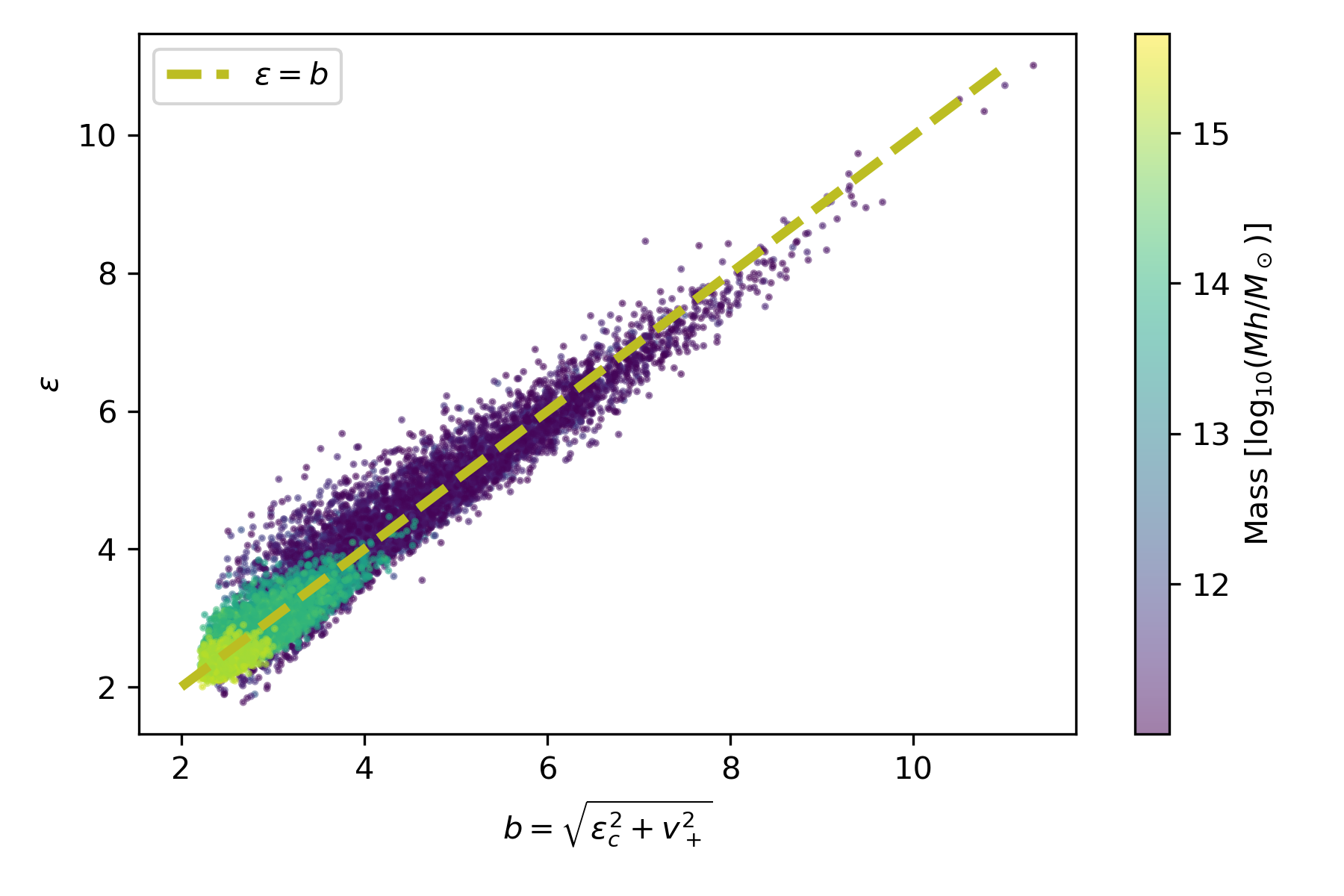}
    \caption{Scatter plot of $\epsilon$ vs $\sqrt{\epsilon_c^2+v_+^2}$, with $v_+=q X(\mu)$ and $\epsilon_c=2.2$, coloured by mass. The residual scatter at fixed $v_+$ is smaller than the range covered as $v_+$ changes.}
    \label{fig:eps_vs_sqrt}
\end{figure}

As we have seen, the scatter is substantially reduced in the $\epsilon$--$v_+$ plane, where $v_+ = qX$. Moreover, the threshold should be such that $\epsilon>v_+$, in order to guarantee the positivity of all eigenvalues and thus triaxial collapse. Therefore, we consider a threshold that depends on $v_+$ rather than $q$.  Noting that there appears to be no additional correlation with the other natural variable $v_-/v_+$, we suppose that progenitors of haloes identified at the present time satisfy 
\begin{equation}
 \epsilon = b + y,\quad {\rm with}\quad 
 b \equiv \sqrt{\epsilon_c^2 + v_+^2} ,
 \label{eq:b+y}
\end{equation}
where $y$ is some other variable whose variance we assume to be small, so that the condition $\epsilon>v_+$ is not violated. 

For example, a contribution to the scatter at fixed $v_+$ (hence, to $y$) would come from differences in the actual virialization time of haloes identified at the same redshift $z=0$ \cite[][]{borzy_colltime14}.  The details of the virialization process will in fact also depend on whether the initial profile is more or less centrally peaked, an effect that is completely independent from the anisotropy of the infall.
The solid line in Figure \ref{fig:eps_vs_uX} shows $\epsilon=b$ (with $\epsilon_c=2.2$), and the dashed one shows the lower limit, $\epsilon=v_+$.
This plot confirms that the distribution of $y$ is narrow compared to that of $b$, suggesting that $v_+$ accounts for most of the measured scatter in $\epsilon$.

To further support our assumption, Figure~\ref{fig:eps_vs_sqrt} shows $\epsilon$ versus $b$, color coded by protohalo mass. The range of $\epsilon$ values at fixed $b$ (the vertical scatter) is significantly smaller than the range spanned as $b$ changes (the diagonal extension of the plot), especially at small and intermediate masses.   This suggests to neglect $y$ in first approximation, adopting
\begin{equation}
    \mathcal{C} 
    = \epsilon- \sqrt{\epsilon_c^2 + v_+^2} 
    = \epsilon - b \simeq 0
  \label{eq:Capprox}
\end{equation}
as the constraint. This assumption reproduces the asymptotic behaviour of the constraint at both large and small $v_+$, and automatically guarantees the positivity of all eigenvalues.

Regardless of its expression, the threshold depends implicitly on mass through the smoothing scale of its variables. Imposing the constraint therefore introduces a mass dependence in all pdfs, which were so far completely mass-independent when expressed in terms of variables normalized by $\sigma_{02}$. In fact, if we write the conditional distribution
\begin{equation}
     p\big(\mathrm{variable|}\mathcal{C}=0\big), 
     \label{eq:pcondC+}
\end{equation}
where the variable could be $\epsilon/\sigma_{02}$, $b/\sigma_{02}$ or $q/\sigma_{02}$, then the constraint will introduce an explicit dependence on mass through $\epsilon_c/\sigma_{02}(M)$.
Comparing these conditional distributions with measurements in our mass-segregated samples will provide further validation of our ansatz on the constraint.

Before moving on, we should note that the most appropriate expression to describe the measurement of a variable in a set of protohaloes of given mass should actually be
\begin{equation} 
  p\big(\mathrm{variable|}\mathcal{C}=0,\mathrm{peak},\mathrm{1st}\big), 
  \label{eq:pC+exact}
\end{equation}
with the conditions including also the peak and first crossing constraint (that is, null gradient, negative definite Hessian, and threshold not having been crossed on any larger scale).  These conditions provide a more accurate description of protohaloes, and at the same time introduce new variables that may also correlate with the stochasticity in $y$ that we are currently ignoring.  However, including these effects is more complicated, and the resulting expressions are not always analytical. For these reasons, we stick to the simpler equation \eqref{eq:pcondC+} in this paper.

\subsection{Constrained $q$ distribution and rescaling of variances}

We have already noticed (histograms of Fig. \ref{fig:hists}, top panel) that the measured distribution of $q/\sigma_{02}$ is nearly mass-independent. We need to verify that this remains true for the predicted distribution in the presence of the mass-dependent constraint equation~(\ref{eq:Capprox}). Since $4q^2=v_+^2+3v_-^2$, the conditional distribution for $q$ is
\begin{equation}
    p\big(q|\mathcal{C}=0\big) 
    = \frac{\avg{\dirac\Big(q-\frac{1}{2}\sqrt{v_+^2+3v_-^2}\Big)\, \dirac(\mathcal{C})}}{\avg{\dirac(\mathcal{C})}}\,,
\end{equation}
where $\avg{\dots}$ means that we average using Equation \eqref{eq:joint_evpvm} (in the numerator) or Equation \eqref{eq:joint_eps_v+} (in the denominator). (Note that the `+' condition is automatically implied by enforcing $\mathcal{C}=0$.)

The denominator of the expression above evaluates to 
\begin{equation}
    \avg{\dirac(\mathcal{C})} = 
    \frac{\mathrm{e}^{-\epsilon_c^2/2}}{36\sqrt{\pi}} 
    \sqrt{5 \left(29-6 \sqrt{6}\right)}\,,
\end{equation}
which, after reintroducing $\sigma_{02}(M)$ with the change $\epsilon_c\to\epsilon_c/\sigma_{02}$, is exponentially mass dependent. However, the exponential cancels out with a similar term in the numerator, and we get
\begin{align}
    p\big(q|\mathcal{C}=0\big) &= 
    \frac{15^2\sqrt{2}q\mathrm{e}^{-5q^2/2}}{\sqrt{\left(29-6 \sqrt{6}\right)}}
    \int_{q}^{2q}\!\dd v \frac{\mathrm{e}^{-v^2/2}}{\sqrt{2\pi}}(v^2-q^2) 
    \nonumber\\
    &=\frac{15^2 q\mathrm{e}^{-5q^2/2}}{\sqrt{\pi\left(29-6 \sqrt{6}\right)}}
    \Bigg[q\bigg(\mathrm{e}^{-q^2/2}-2 \mathrm{e}^{-2 q^2}\bigg)\nonumber \\
    &\quad + \!\!\left.\sqrt{\frac{\pi}{2}}\left(q^2-1\right) 
    \left(\text{erf}\left(\frac{q}{\sqrt{2}}\right)-\text{erf}\left(\!\sqrt{2} q\right)\right)
    \right],
    \label{eq:pqC+}
\end{align}
where there is no dependence on $\epsilon_c$, and thus no mass dependence left. In fact, this expression is numerically quite close to equation \eqref{eq:pdfq+}; in particular, it is closer to the dashed curve in Fig. \ref{fig:hists} than to the histograms there.  In other words, using a better model for the threshold does not fix the disagreement with the measured distribution of $q$ for protohaloes.  

\begin{figure}
    \centering
    \includegraphics[width=\columnwidth]{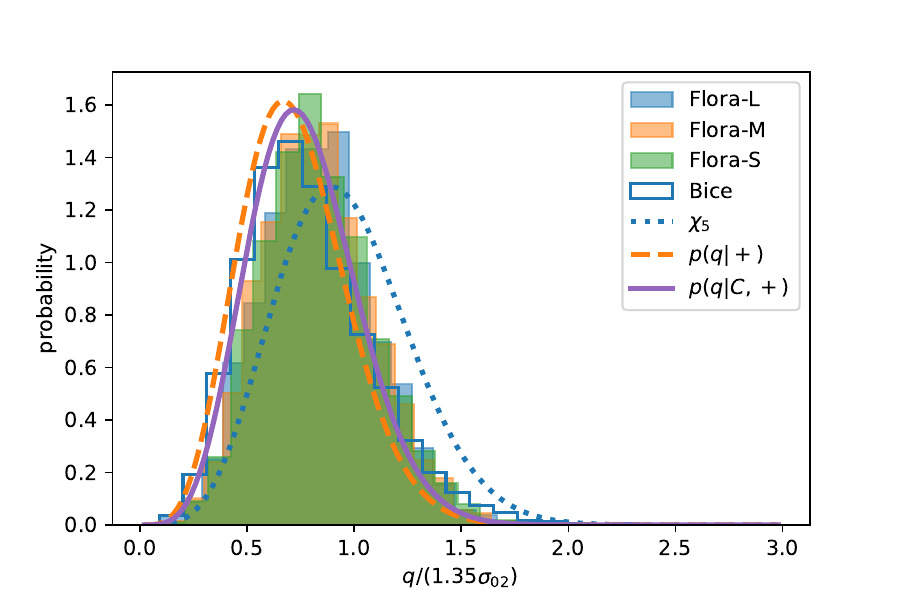}
    \caption{Distribution of $q/(1.35\sigma_{02})$, the rescaling that provides the best match to equation \eqref{eq:pqC+} (purple solid line). Histograms show the measurements; dotted, dashed and solid curves show the distribution of random positions, those constrained to be positive definite, and those constrained to also have $\epsilon=b$.}
    \label{fig:hist_qu_rescaled}
\end{figure}

We should not expect to find perfect agreement, though, since protohalo patches are not spherical, whereas our analysis has implicitly assumed they are. This will for instance change the prediction of the variance. 
To account at least qualitatively for this, we rescale $\sigma_{02}$ (the standard deviation of $\epsilon$ in spheres containing a given mass) by an ad-hoc factor whose value we motivate elsewhere \citep[see][for earlier work]{paddyKandu93,brlc2002,ls08}. 
Here, we simply note that this factor should be greater than unity, to compensate for the fact that $\epsilon$ in the protohalo shapes is always larger than in the corresponding Lagrangian sphere \citep{eshape23}. We also assume for simplicity that this factor is the same for all masses. To fix its amplitude, we ask that the distribution of $q/\sigma_\epsilon$  match equation \eqref{eq:pqC+} closely.  Fig. \ref{fig:hist_qu_rescaled} shows that setting $\sigma_\epsilon= 1.35\,\sigma_{02}$ provides reasonably good agreement:  the rescaled histograms are closer to the solid curve than to the dotted.  Presumably, some of the small differences that remain are due to approximating equation~(\ref{eq:pC+exact}) with equation~(\ref{eq:pcondC+}), and ignoring the stochasticity which allowed us to approximate equation~(\ref{eq:b+y}) with equation~(\ref{eq:Capprox}).

\subsection{Testing the threshold ansatz}

We are now going to test if equation~(\ref{eq:Capprox}) is a useful parametrization of the threshold for collapse, by comparing the measured distributions of $\epsilon/\sigma_{02}$ and $b$ with the predicted conditional distribution. The latter can be expressed as
\begin{equation}
    p(b|\mathcal{C}=0) = \frac{\int_0^b\dd v_+\delta_\mathrm{D}(b-\sqrt{\epsilon_c^2+v_+^2})p(\epsilon=b,v_+)}
    {\int_0^\infty\dd\epsilon\int_0^\epsilon\dd v_+\delta_\mathrm{D}(\epsilon-\sqrt{\epsilon_c^2+v_+^2})p(\epsilon,v_+)}\,;
    \label{eq:pbdef}
\end{equation}
Explicit calculation, using $p(\epsilon,v_+)$ from equation \eqref{eq:joint_eps_v+}, yields
\begin{equation}
    p(b|\mathcal{C}=0) = \frac{18b\,{\rm e}^{-9v_+^2/8}}{\sqrt{\pi v_+^2}}\,
    \frac{{\rm e}^{-15v_+^2/8} -1 + 15v_+^2/8}{\sqrt{29 - 6\sqrt{6}}}
    \bigg|_{v_+^2=b^2-\epsilon_c^2}
\label{eq:pb}
\end{equation}
where $2b\,db = 2v_+\,dv_+$ is the reason for the $b/v_+$ factor in the expression above. Of course, this assumes that $b\ge\epsilon_c$, otherwise $p(b|\mathcal{C}=0)=0$.  

This distribution explicitly depends on $\epsilon_c$, that is on $\epsilon_c/\sigma_{02}$ after reintroducing $\sigma_{02}$, so we must be careful when comparing it with measurements that span a range of masses.  (This was not a concern for $q/\sigma_{02}$ for which there was no predicted mass-dependence.)  
To make a fair comparison, we must average equation \eqref{eq:pb} over the same set of masses that contributed to the measurement.  Since $\sigma_{02}(M)$ is monotonic, this average is   
\begin{equation}
   p(b|{\rm range}) 
= \frac{\int d\sigma_{02}\,n(\sigma_{02})\,p(b|\mathcal{C}=0)}
       {\int d\sigma_{02}\,n(\sigma_{02})}.
\end{equation}
Ideally, we would like to {\em predict} $n(\sigma_{02})$, rather than use the measured distribution.  While a more careful derivation is the subject of work in progress, and we provide a simple estimate, motivated by a Press-Schechter like approach, in Appendix \ref{sec:app_PS}, the plots to follow were made by summing over the actual distribution of $\sigma_{02}$ values measured in each sample.  

\begin{figure}
    \centering
    \includegraphics[width=\linewidth]{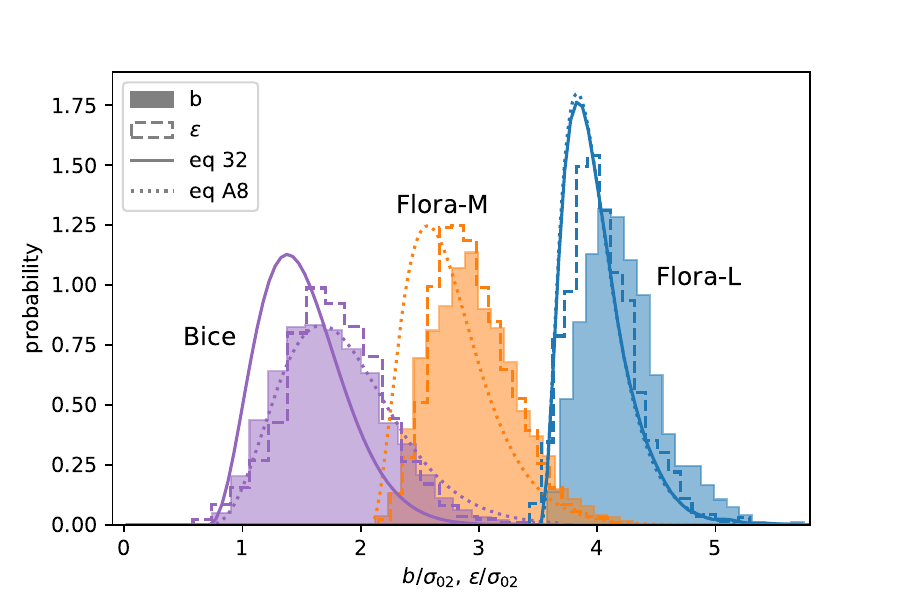}
    \caption{Distribution of $b = \sqrt{\epsilon_c^2+v_+^2}/\sigma_{02}$ in the four mass bins (filled histograms). Dashed lines show the  distribution of $\epsilon/\sigma_{02}$ (from Figure~\ref{fig:hist_eps}); solid lines show equation \eqref{eq:pb} averaged over the measured distribution of $\sigma_{02}$ values; dotted lines show equation~(\ref{eq:pb_range}).}
    \label{fig:hist_sqrt_barr}
\end{figure}

With this in mind, we now proceed to a quantitative comparison. The filled histograms of Fig. \ref{fig:hist_sqrt_barr} show the distribution of $b$ in three of our protohalo samples, and the dashed step lines the measured distribution of $\epsilon$ (i.e. the same histograms as in Fig. \ref{fig:hist_eps}).  
The similarity of the two sets of histograms proves that setting $\epsilon^2 = \epsilon_c^2+v_+^2$ accounts for most of the scatter and is therefore a very good approximation for the threshold for halo formation.  

\begin{figure}
    \centering
    \includegraphics[width=\columnwidth]{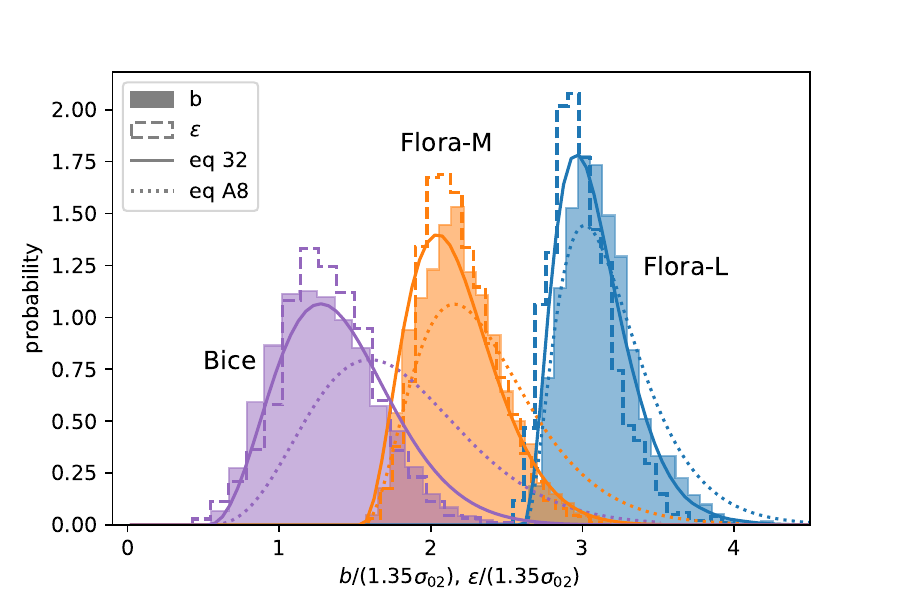}
    \caption{Weighted average of equation \eqref{eq:pb} (solid curves) with $\omega_c=\epsilon_c/(1.35\sigma_{02})$, as described in the main text, providing an excellent description of the rescaled histograms of $b/(1.35\sigma_{02})$ for all samples.  Dotted curves show \eqref{eq:pb_range}.}
    \label{fig:hist_sqrt_barr_rescaled}
\end{figure}

\begin{figure}
    \centering
    \includegraphics[width=\columnwidth]{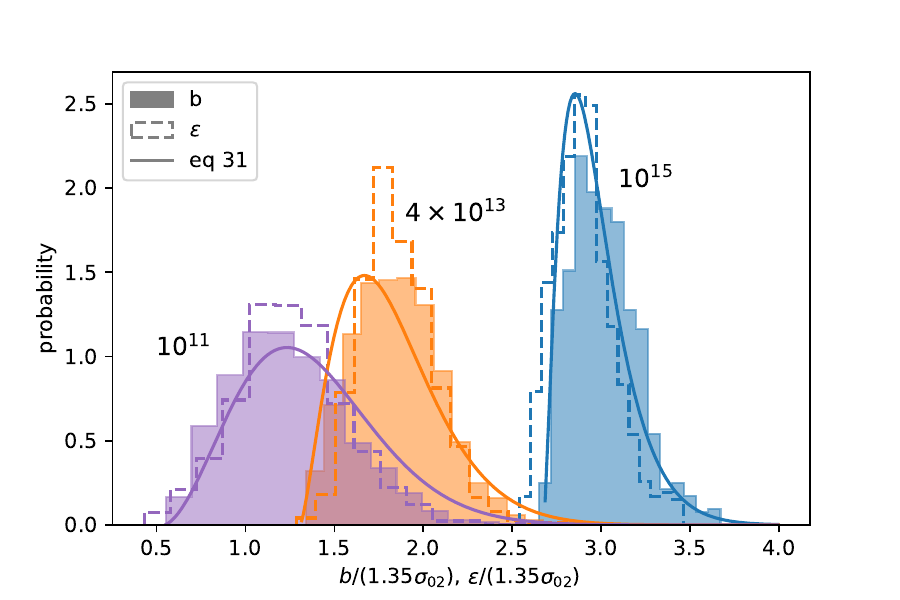}
    \caption{Distribution of $b/(1.35\sigma_{02})$ (filled histograms) and $\epsilon/(1.35\sigma_{02})$ (empty dashed) in three narrow mass bins of mass  $10^{11}$, $4\times10^{13}$ and $10^{15}h^{-1}M_\odot$ (from left to right). Solid curves show equation~\eqref{eq:pb}.}
    \label{fig:narrowbins}
\end{figure}

Notice that the measured histograms are shifted to larger values of $b/\sigma_{02}$ (compared to the theory curves).  In our discussion of $q$, we noted that rescaling $\sigma_{02}\to 1.35\sigma_{02}$ worked quite well, so it is natural to ask if this also works well here.  Fig.~\ref{fig:hist_sqrt_barr_rescaled} shows that it does.  
As further confirmation of the accuracy of this rescaling,  Fig. \ref{fig:narrowbins} compares the distribution of $b/1.35$  measured in narrow mass bins directly with equation \eqref{eq:pb}. We choose three mass intervals of (1 -- 2)$\times10^{11}$, (4 -- 5)$\times10^{13}$ and (1-1.25)$\times10^{15}h^{-1}M_\odot$, corresponding to the lower end of Bice, Flora-S and Flora-L respectively. The bins are sufficiently narrow that no averaging over $\sigma_{02}$ values should be needed.  Therefore, for the theory curves, we use as $\omega_c$ the mean of $\epsilon_c/(1.35\sigma_{02})$ in each narrow bin.  

The remarkable agreement with the filled histograms confirms that simply requiring the positivity of the eigenvalues, and imposing the approximate constraint $\epsilon^2 = \epsilon_c^2+v_+^2$, gets very close to describing the correct distribution of $v_+$ (and thus of $b$) in protohaloes. In turn, $b$ is a rather good approximation to the value of $\epsilon$, and therefore to the collapse threshold of protohaloes.
While a more careful calculation would require that we also impose the excursion set and peaks constraints (see equation~\ref{eq:pC+exact} and related discussion), we believe that the considerably simpler calculation discussed here (equations~\ref{eq:pcondC+} and~\ref{eq:Capprox}) captures most of the interesting physics.

\section{Discussion and conclusions} 

We have explored the statistical implications of two facts of dynamical origin that make protohaloes---the initial patches that are destined to become haloes---special locations: they must collapse along three axes, and they have to do so by today. 
Expressed in mathematical terms, these conditions are:  
\begin{itemize}
 \item[(a)] the energy shear $u_{ij}$ is positive definite, so that all three axes of the protohalo collapse \citep{mds24};    
 \item[(b)] $\epsilon$, the trace of $u_{ij}$, must exceed a critical value for the collapse to complete by the present day \citep{epeaks}.
\end{itemize}
Both conditions are interesting in their own right.  

The energy shear tensor $u_{ij}$ (equation~\ref{eq:uijdef}) measured at random positions in a Gaussian random field is a Gaussian random matrix, and its three rotational invariants (equation~\ref{eq:eumu}) are therefore uncorrelated with one another.  However, in protohaloes they are strongly correlated (Fig.~\ref{fig:eps_vs_u}).  
We have shown that the measured correlation pattern---in particular, the limiting values of some of these correlations (Figs.~\ref{fig:eps_vs_u} and~\ref{fig:eps_over_mu})---follows directly from the positive definiteness of the energy shear, that is from Condition (a).  We provided analytic expressions for the distributions of the three invariants, and their correlations, subject to these conditions (equations~\ref{eq:pdfe+}--\ref{eq:pdfmu+}, and~\ref{eq:pdfe+2}--\ref{eq:pdfmu+2} if one also requests $\epsilon>\epsilon_c$).  

Previous work has shown that the collapse threshold in $\epsilon$ is stochastic, with an absolute lower bound that we denote $\epsilon_c$.  
Fig.~\ref{fig:eps_vs_u} shows that, for the protohaloes of haloes identified at $z=0$, 
$\epsilon_c\simeq 2$.  
Although Condition (a) cannot explain the lower bound, the positivity requirement means that the excursion above this limit actually strongly correlates with the other rotational invariants (built out of differences of the eigenvalues). We have verified that a parametrization of the stochasticity based on $v_+$ (equation~\ref{eq:vpm}) appears to be more efficient--- that is, it has significantly less residual scatter---than one based on the rotational invariants $q$ and $\mu$ (compare Figs.~\ref{fig:eps_vs_u} and Figs.~\ref{fig:eps_vs_uX}).
In particular, $\epsilon$ in protohaloes is very close to $b=\sqrt{\epsilon_c^2 + v_+^2}$ (Fig.~\ref{fig:eps_vs_sqrt}). Parametrising the threshold with this expression accounts for most of its stochasticity.

We have then computed the conditional distribution of $q$ given that $\epsilon$ equals the threshold. The result is rather insensitive to the constraint, and it matches the measured distributions well provided that one rescales $\sigma_{02}$ (the standard deviation of $\epsilon$ in spherical volumes) by a factor of 1.35 (Fig. \ref{fig:hist_qu_rescaled}). We argued that this rescaling can be attributed to the fact that protohaloes are not spherical. Computing it from first principles is the object of current work.  
The conditional distribution of $b$ (equation~\ref{eq:pb}) is on the other hand very sensitive to the constraint, which makes it strongly mass dependent. However, the same variance rescaling brings the measurements of this quantity into good agreement with our (mass-dependent) predictions.

To summarize, we began our study with two main goals.  One was the question of whether $\mu$, the third order invariant of the energy shear tensor, matters for protohaloes.  Fig.~\ref{fig:eps_vs_u} shows that it does, which has therefore potential implications for assembly bias. Our subsequent analysis suggests that a more efficient parametrisation of the physics of collapse uses $v_+$,  a nonlinear combination of the second and third order invariants ($q$ and $\mu$) that is actually linear in the eigenvalues themselves (equation~\ref{eq:vqX}). 

The second goal was to determine if the correlations between the invariants of the energy shear measured at protohalo positions (at random positions they are independent) can be understood as simply arising from the positivity of the eigenvalues.  Here too the answer is `yes', but with some subtleties.  
Our analysis suggests that triaxial collapse requires $\lambda_3\ge 0$, but collapse by today means $\epsilon\ge 2$. For very massive protohaloes the dominant condition is the latter: in this regime, $\epsilon\sim 2$ with very little residual dependence on other rotational invariants. For smaller objects, however, which typically have $\epsilon\gg 2$, much of the correlation between $\epsilon$ and $q$ or $v_+$ emerges from the constrained statistics, rather than being a new physical requirement.  The positivity condition is much simpler in the basis $(\epsilon,v_+,v_-)$ than in $(\epsilon,q,\mu)$, which explains why parametrising the stochasticity of the threshold with $v_+$ is more efficient.  

Our predictions are completely analytical and reasonably simple. This contrasts with many results on protohalo statistics---typically based on peak theory and first passage probabilities---which require rather involved calculations. These complications are indeed necessary when computing number densities, masses and correlation functions. However, a byproduct of our analysis is that they are not always required to predict the probability of secondary quantities in protohaloes. For these, conditioning on a sufficiently good model of the threshold is enough.

These results bring us closer to a first principles derivation of the abundance of dark matter haloes and their properties.  The three specific remaining issues that still need to be addressed are:
\begin{enumerate}
 \item[(i)] the impact of non-sphericity of protohalo shapes on their statistics; 
 \item[(ii)] the impact on the statistics of accounting for the fact that protohalo positions define a point process; and   
 \item[(iii)] the exact relation between the value of $\epsilon$ and the time of virialisation.
\end{enumerate}
About the first point, our work suggests that the main effect of initial non-sphericity is to rescale all variances:  in a follow-up to \cite{eshape23}---requiring protohalo shapes to maximise $\epsilon$---we are in the process of showing why. In addition, now that we have a good model for the critical threshold (equation~\ref{eq:b+y}), we believe we are in a position to apply the technology of excursion set energy peaks theory \citep{epeaks} to address point (ii).  This will allow us to quantify assembly or secondary bias effects as well \citep{st04,dwbs08,ms13b,lms17,bvTaka,fnlAB}. It might also help improving models of shape bias for intrinsic alignments \citep{Blazek_shapebias2015,Vlah_shapebias2020,Lee_shapebias2024,jens_shapebias2024}

As for point (iii), we should note that a halo identified at $z=0$ may have virialised at an earlier time: halo finders do not usually make this distinction. Since the evolution of the inertial radius follows in first approximation spherical collapse, where the final time  depends on the initial energy, uncertainty in the virialisation time will add to the stochasticity in the value of $\epsilon$ \citep[similarly to][]{borzy_colltime14}.
Much of this stochasticity, as shown in this work, is described by $v_+$: when $v_+$ is large then so is $\epsilon$, implying early virialisation. Moreover, $\epsilon\sim v_+$ means that $\lambda_3\sim 0$: such objects represent the transition between a protohalo and a protofilament \citep{Hanami,critev,CUSP2021}, and can be associated to haloes that stopped accreting before the time of identification due to tidal stripping \citep[as found in simulations by][]{ZOMG-I2017}. There is however also a residual scatter in $\epsilon$ at fixed $v_+$, which might be related to how virialisation depends on the slope of the initial profile. We hope to quantify this effect in future work.

To get a more reliable estimate of the virialisation time in an anisotropic environment (large $v_+$), one may consider perturbative corrections to spherical collapse. These  corrections will naturally scale like $q^2/\epsilon^2$. Our analysis suggests that this ratio cannot exceed unity, which helps keep the perturbative treatment under control.  Physically, this can be interpreted as reflecting the fact that the evolution cannot deviate too much from spherical collapse, even for arbitrary shapes, simply because the protohalo patch must end up in a very small region in order to create a large enough density. This is what makes protohaloes special locations.

Finally, our goal has been to predict where haloes form in the {\em initial} field.  Recent machine-learning based work aims to predict where they occur in the {\em evolved} field \citep{gotham}.  We hope that our work will be useful for training such algorithms to make predictions from the initial field.

\section*{Acknowledgements}

Thanks to G. Despali for sharing the data from the SBARBINE simulations,
and to the organisers and participants of the 2023 KITP Cosmic Web workshop. Our visit to KITP was supported by the National Science Foundation under PHY-1748958.  We are grateful to the EAIFR, Kigali and the ICTP, Trieste for their hospitality in summer 2024, when this work was completed.   


\section*{Data Availability}

The simulation data and the post-processing quantities used in this work can be shared on reasonable request to the authors.
 

%
\bibliographystyle{aa} 
\bibliography{mybib} 
%

\begin{appendix}
\section{Statistics for a range of masses}
We discuss a simple model for the statistics when the sample includes objects more massive than a given threshold.  The approach is similar to that of \cite{ps74}, which is known to not be very accurate---in particular, it does not account for the fact that haloes define a point process.  Nevertheless, we believe it provides some insight.  

\subsection{Simple threshold:  $\epsilon=\epsilon_c$}
\label{sec:app_ec}
Integrating over masses above a threshold is like integrating $\sigma_{02}$ from 0 to an upper limit.  Since the statistics depend on $\epsilon_c/\sigma_{02}$, this is like integrating $\epsilon_c$ from a lower limit to infinity.   

Hence, as a warm up, consider simply requiring $\epsilon \ge \epsilon_c$.  To quantify how the distribution of $\epsilon$, $q$ and $\mu$ are modified, we first define the new normalisation 
\begin{align}    
    p(+\epsilon_c) \equiv p(\lambda_3\leq0,\epsilon>\epsilon_c) = p(+)\int_{\epsilon_c}^\infty d\epsilon\,p(\epsilon|+) \,.
\end{align}
Although no analytical expression for $p(+\epsilon_c)$ exists, a simple and reasonably accurate approximation is
\begin{align}
    p(+\epsilon_c) &\simeq \frac{1}{4}\mathrm{erfc}\left(\frac{\epsilon_c}{\sqrt{2}\sigma_{02}}\right)
    \left[1+\mathrm{erf}\left(\frac{\epsilon_c}{\sqrt{2}\sigma_{02}}\right)\right] \nonumber\\
    &\quad -\frac{\sqrt{5}}{3\pi}\mathrm{e}^{-9\epsilon_c^2/8\sigma_{02}^2} 
    +\frac{\pi-2\arctan(\sqrt{5})}{4\pi}\mathrm{e}^{-3\epsilon_c^2/2\sigma_{02}^2},
\end{align}
which captures the correct asymptotic behaviour for small and large $\sigma_{02}$ (large and small masses) and deviates from the numerical solution by less than 1.8\% over the entire positive axis.

Equations \eqref{eq:pdfe+}, \eqref{eq:pdfq+}  and \eqref{eq:pdfmu+} are changed to 
\begin{align}
    p(\epsilon|+\epsilon_c) &= \Theta(\epsilon-\epsilon_c)p(\epsilon|+)\frac{p(+)}{p(+\epsilon_c)}\,, 
    \label{eq:pdfe+2}\\
    p(q|+\epsilon_c) &= \frac{\chi(q)}{p(+\epsilon_c)} 
    \left[\int_{q_{\rm min2}}^\infty d\epsilon\,p(\epsilon)\right. \nonumber\\
    &\quad  
        \left.+\int_{q_{\rm min1}}^{q_{\rm min2}} d\epsilon\,p(\epsilon)\,\frac{(\epsilon/2q)(\epsilon^2/q^2 - 3) + 1}{2}\right] \nonumber\\
    &= \frac{\chi(q)}{4p(+\epsilon_c)}\Big[{\rm erfc}\left(\frac{q_{\rm min1}}{\sqrt{2}}\right) + {\rm erfc}\left(\frac{q_{\rm min2}}{\sqrt{2}}\right) \nonumber\\
    &\qquad\qquad\qquad - \frac{{\rm e}^{-q_{\rm min2}^2/2}}{\sqrt{2\pi}}\frac{(q_{\rm min2}^2 + 2 - 3q^2)}{q^3} \nonumber\\
    &\qquad\qquad\qquad +  
    \frac{{\rm e}^{-q_{\rm min1}^2/2}}{\sqrt{2\pi}}
    \frac{(q_{\rm min1}^2 +2 - 3q^2)}{q^3} \,
                 \Big] 
                 \label{eq:pdfq+2}
\end{align}
where $q_{{\rm min}n}\equiv {\rm max}(\epsilon_c/\sigma_{02},nq)$, and
\begin{align}
   p(\mu|+\epsilon_c) &=  \frac{1}{2p(+\epsilon_c)} \int_{\epsilon_c}^\infty\dd\epsilon p(\epsilon) \int_0^{\epsilon/X}\dd q \chi(q)  \nonumber \\
   & = \int_{\epsilon_c}^\infty\dd\epsilon \frac{p(\epsilon)}{2p(+\epsilon_c)}\bigg[ \text{erf}\left(\sqrt{\frac{5}{2}} \frac{\epsilon }{X}\right) \nonumber \\
   &\qquad\qquad\qquad -\sqrt{\frac{10}{\pi }} \epsilon  \frac{ 5 \epsilon ^2+3 X^2}{3 X^3}\mathrm{e}^{-5 \epsilon ^2/2 X^2}\bigg]\,.
   \label{eq:pdfmu+2}
\end{align}
The last integral does not have an analytical expression, but an approximation accurate to better than 1.5\% is
\begin{align}
    p(\mu|+\epsilon_c) &\simeq \frac{1}{p(+\epsilon_c)}\bigg[
    \frac{1}{8} \left(\text{erf}\left(\frac{\omega _c}{\sqrt{2}}\right)+1\right) \text{erfc}\left(\frac{\omega _c}{\sqrt{2}}\right) \nonumber \\
    &\qquad +\left(\frac{\arctan\left(\sqrt{5}/X\right)}{2 \pi }-\frac{1}{8}\right) \mathrm{e}^{-\left(X^2+5\right) \omega _c^2/4 X}\bigg]\,.
\end{align}

Massive objects correspond to the $\epsilon_c/\sigma_{02}\gg 1$ limit, for which $p(+\epsilon_c) \simeq \frac{1}{2}\mathrm{erfc}(\epsilon_c/\sqrt{2}\sigma_{02})$ and $q_{{\rm min}1}=q_{{\rm min}2}=\epsilon_c/\sigma_{02}$. Therefore, only the first term of equation \eqref{eq:pdfq+} contributes and tends to $\chi(q)$ of Eq.~(\ref{eq:chi5}) (dotted curve in Fig.~\ref{fig:hists}).  In contrast, at small masses the limits of the integrals in Eq.~(\ref{eq:pdfq+}) are modified only when $q\le 2$ (i.e. not significantly).  As a result, lower mass haloes are better described by Eq.~(\ref{eq:pdfq+}) (dashed curve in Fig.~\ref{fig:hists}).  

These limiting cases can also be understood by looking at Fig.~\ref{fig:eps_vs_u} which shows the two cuts explicitly: the dashed line showing $\epsilon=q$ and the obvious floor at $\epsilon=2$.  Most lower mass haloes tend to have large $q$, so they are less affected by the $\epsilon>2$ floor: the $\epsilon>q$ cut matters much more.  However, massive haloes tend to have small $q$, so the $\epsilon>2$ floor is the only relevant cut.  In a Gaussian field, the distribution of $q$ is uncorrelated with $\epsilon$.  This is modified by the $\epsilon=q$ cut, but, since this cut is irrelevant for the massive haloes, their distribution is not given by Eq.~(\ref{eq:pdfq+}); instead, it should be much closer to Eq.~(\ref{eq:chi5}).  Fig.~\ref{fig:hists} shows that the distribution of massive haloes is indeed shifted towards larger values, as expected.  

By this argument, one would expect the distribution of $\mu$ for massive haloes to be closer to that for unconstrained positions (uniform on $[-1,1]$); this is in qualitative agreement with the trend in the bottom panel of Fig.~\ref{fig:hists}, where the largest mass objects (green) are closest to being uniformly distributed.

\subsection{More realistic threshold:  $\epsilon = \sqrt{\epsilon_c^2 + v_+^2}$}
\label{sec:app_PS}
Now suppose the threshold is given by equation~(\ref{eq:b+y}) of the main text (and the extra variable $y$ is ignored).  Integrating over all masses above some minimum means we now want $\sqrt{\epsilon^2-v_+^2}/\sigma_{02}>\omega_c$, or equivalently $v_+/\sigma_{02}<\sqrt{\omega^2-\omega_c^2}$. The resulting distribution is then
\begin{equation}
    p(b|\mathcal{C}>0) \propto 
    \int_0^{\sqrt{b^2-\omega_c^2}}\dd v_+p(\epsilon=b,v_+)\,,
    \label{eq:pb_range_def}
\end{equation}
where the integral is analytical and is easily seen to be equal to ${\rm e}^{-\omega_c^2/2}p(\sqrt{b^2-\omega_c^2}|+)p(+)$. Normalising to unity, we get
\begin{equation}
    p(b|\mathcal{C}>0) = 
    \frac{p(\epsilon=\sqrt{b^2-\omega_c^2}|+)}{N(\omega_c)}\,,
    \label{eq:pb_range}
\end{equation}
where $N(\omega_c)$ is the integral of the numerator over $b>\omega_c$. Changing variables from $b$ to $\sqrt{b^2-\omega_c^2}$, the normalisation becomes
\begin{align}
  N(\omega_c) = \int_0^{+\infty}\dd\epsilon
  \frac{\epsilon}{\sqrt{\epsilon^2+\omega_c^2}} p(\epsilon|+)\,;
\end{align}
this integral has no known analytical solution, but can be approximated by Taylor-expanding the fraction $\epsilon/\sqrt{\epsilon^2+\omega_c^2}$ around the conditional mean $\avg{\epsilon|+}\simeq1.663$. At second order, this gives
\begin{equation}
    N(\omega_c) \simeq 
    \frac{\avg{\epsilon|+}}{\sqrt{\avg{\epsilon|+}^2+\omega_c^2}}
    \bigg[1-\frac{3\mathrm{Var}(\epsilon|+)}{(\avg{\epsilon|+}^2+\omega_c^2)^2}
    \frac{\omega_c^2}{2}\bigg] ,
    \label{eq:Nbetter}
\end{equation}
where $\mathrm{Var}(\epsilon|+)\simeq0.3010$. This approximation is asymptotically correct for both large and small $\omega_c$ (in which limits one gets $\avg{\epsilon|+}/\omega_c$ and 1 respectively), and numerically accurate to less than 0.4\% everywhere. 
The dotted curves in Fig. \ref{fig:hist_sqrt_barr} show equation \eqref{eq:pb_range}.  While the agreement with the measurements is not compelling, as we noted above, this is only a crude estimate.  
A better estimate would include the peak and first-crossing constraints, and the corresponding Jacobians: the Hessian determinant and the slope of the excursion set \citep{epeaks}. This is the subject of work in progess.

\end{appendix}

\end{document}